\documentclass[11pt, a4paper]{article}
\pdfoutput=1
\usepackage{graphicx}
\usepackage{amssymb}
\usepackage{amsmath}
\usepackage{bm}
\usepackage{url}
\usepackage[table]{xcolor} 
\usepackage{cite}
\usepackage{slashed}
\usepackage{subfigure}
\usepackage{epstopdf}            
\usepackage{epsfig}
\usepackage{here}
\usepackage{comment}
\usepackage{xfrac} 
\usepackage{multirow}
\usepackage{adjustbox}
\usepackage{booktabs} 
\usepackage{colortbl} 

\renewcommand{\thefootnote}{\fnsymbol{footnote}}
\numberwithin{equation}{section} 
\def\beq#1\eeq{\begin{align}#1\end{align}}

\renewcommand{\arraystretch}{1.3}

\newcommand{\eg}{{\em e.g.}}
\newcommand{\ie}{{\em i.e.}}

\RequirePackage{xspace}
\def\Bbar    {\kern 0.18em\overline{\kern -0.18em B}{}\xspace}
\def\Bb      {\ensuremath{\Bbar}\xspace}

\def\Jpsi   {J\mbox{\footnotesize$\!/\psi$}}

\usepackage{caption}
\definecolor{BlueViolet}{rgb}{0.2, 0.00, 0.7}
\definecolor{Blue}{rgb}{0.15, 0.00, 0.9}
\definecolor{light_blue}{rgb}{0.15, 0.35, 0.95}
\definecolor{kit_green}{rgb}{0, 
0.58823 
, 0.50980 
}
\usepackage[
colorlinks=true, linkcolor=light_blue,citecolor=light_blue,urlcolor=kit_green]{hyperref} 
\usepackage[hyphenbreaks]{breakurl} 

\normalsize 

\begin{document}
\sloppy 

\begin{titlepage}

\begin{center}
\begin{flushright}
P3H--22--103, TTP22--062, KEK--TH--2464\\ 
\end{flushright}

\vskip .5in

{\fontsize{19pt}{0cm}\selectfont
 {\bf  Global fit to 
$\boldsymbol{b \to c\tau\nu}$ anomalies
2022 mid-autumn
}
}

\vskip .2in

{\large 
{\bf Syuhei Iguro,$^{\rm (a,b)}$
Teppei Kitahara,$^{\rm (c,d,e,f)}$
and 
Ryoutaro Watanabe$^{\rm (g)}$}}

\vskip .3in

\begingroup\small
\begin{tabbing}
$^{\rm (a)}$ \= {\it 
Institute for Theoretical Particle Physics (TTP), Karlsruhe Institute of Technology (KIT),}\\
\> {\it Engesserstra{\ss}e 7, 76131 Karlsruhe, Germany}
\\[0.2em]
$^{\rm (b)}$ \> {\it Institute for Astroparticle Physics (IAP),
KIT, 
Hermann-von-Helmholtz-Platz 1,}\\
\> {\it 76344 Eggenstein-Leopoldshafen, Germany}
\\[0.2em]
$^{\rm (c)}$ \> {\it 
Institute for Advanced Research, Nagoya University, Nagoya 464--8601, Japan}
\\[0.2em]
$^{\rm (d)}$ \> {\it 
Kobayashi-Maskawa Institute for the Origin of Particles and the Universe, Nagoya University, } \\
\> {\it Nagoya 464--8602, Japan}
\\[0.2em]
$^{\rm (e)}$ \> {\it 
KEK Theory Center, IPNS, KEK, Tsukuba 305--0801, Japan} 
\\[0.2em]
$^{\rm (f)}$ \> {\it 
CAS Key Laboratory of Theoretical Physics, Institute of Theoretical Physics, } \\
\> {\it Chinese Academy of Sciences, Beijing 100190, China}
\\[0.2em]
$^{\rm (g)}$ \> {\it 
INFN, Sezione di Pisa, Largo B. Pontecorvo 3, 56127 Pisa, Italy}
\end{tabbing}
{\href{mailto:igurosyuhei@gmail.com}{igurosyuhei@gmail.com}, \href{mailto:teppeik@kmi.nagoya-u.ac.jp}{teppeik@kmi.nagoya-u.ac.jp}, \href{mailto:wryou1985@gmail.com}{wryou1985@gmail.com}}\\
\endgroup 
\vskip 0.05in
{Dated: October~18, 2022} 
\end{center}
\vskip .1in

\begin{abstract}
\noindent 
Recently, the LHCb collaboration announced a preliminary result of the test of lepton flavor universality (LFU) in $B\to D^{(\ast)}$ semi-leptonic decays: 
$R_{D}^{\rm LHCb2022} = 0.441 \pm 0.089$ and 
$R_{D^{\ast}}^{\rm  LHCb2022} = 0.281 \pm 0.030$ based on the LHC Run\,1 data.
This is the first result of $R_{D}$ for the LHCb experiment, and its precision is comparable to the other $B$-factory data.
Interestingly, those data prefer the violation of the LFU again.
A new world average of the data from the BaBar, Belle, and LHCb collaborations is 
$R_{D} = 0.358 \pm 0.027$ and
$R_{D^{\ast}} = 0.285 \pm 0.013$. 
Including this new data, we update a circumstance of the $b \to c \tau \overline\nu$ measurements and their implications for new physics.
Incorporating recent developments for the $B \to D^{(\ast)}$ form factors in the Standard Model (SM),
we observe a $4.1 \sigma$ deviation from the SM predictions. 
Our updates also include; 
model-independent new physics (NP) formulae for the related observables; 
and the global fittings of parameters for leptoquark scenarios as well as single NP operator scenarios. 
Furthermore, we show future potential to indirectly distinguish different new physics scenarios 
with the use of the precise measurements of the polarization observables in $B\to D^{(\ast)}\tau \overline\nu$ at the Belle~II and 
the high-$p_{\rm T}$ flavored-tail searches at the LHC. 
We also discuss an impact on the LFU violation in $\Upsilon \to l^+ l^-$.

\end{abstract}
\end{titlepage}

\setcounter{page}{1}
\renewcommand{\thefootnote}{\#\arabic{footnote}}
\setcounter{footnote}{0}

\hrule
\tableofcontents
\vskip .2in
\hrule
\vskip .4in


\section{Introduction}
\label{sec:intro}

The semi-tauonic $B$-meson decays, $\Bb \to D^{(\ast)} \tau \overline{\nu}$, have been intriguing processes to measure the lepton flavor universality (LFU):
 \begin{align}
  R_D \equiv \frac{\mathcal{B}(\Bb\rightarrow D \,\tau\, \overline\nu_\tau)}{\mathcal{B}(\Bb\rightarrow D\, \ell\,\overline\nu_\ell)} \,, \qquad
  R_{D^{\ast}} \equiv \frac{\mathcal{B}(\Bb\rightarrow D^{\ast} \tau \,\overline\nu_\tau)}{\mathcal{B}(\Bb\rightarrow D^{\ast} \ell\,\overline\nu_\ell)} \,,
\end{align}
since it has been reported that the measurements by 
the BaBar~\cite{Lees:2012xj,Lees:2013uzd}, Belle~\cite{Huschle:2015rga,Hirose:2016wfn,Hirose:2017dxl,Abdesselam:2019dgh,Belle:2019rba} and LHCb~\cite{Aaij:2015yra,Aaij:2017uff,Aaij:2017deq} collaborations 
indicate deviations from the Standard Model (SM) predictions, where $\ell =e,~\mu$ for the BaBar/Belle and $\ell =\mu$ for the LHCb. 
See Table~\ref{tab:RD_exps} for the present summary.

A key feature of the deviation is that the measured $R_{D^{(\ast)}}$ are always {\em excesses} compared with the SM predictions and thus imply violation of the LFU. 
Then it has been followed by a ton of theoretical studies to understand its implication from various points of view, \eg, see Ref.~\cite{London:2021lfn} and references therein. 
A confirmation of the LFU violation will provide an evidence of new physics (NP).
%

\subsection{Summary of the current status: 2022 mid-autumn}

Three years have passed since the previous experimental report of $R_{D^{(\ast)}}$ measurements from the $B$ factories \cite{Belle:2019rba}.
In the meantime, the Belle~II experiment finally started taking data from 2020~\cite{Belle-II:2020sdf,Belle-II:2022cgf},
and the CMS collaboration has developed an innovative data recording method, called ``$B$ Parking'' since 2019 \cite{Bparking,CMS-DP-2019-043,Bainbridge:2020pgi,Takahashi}, 
although their official first results are still being awaited.

On the other hand, the LHCb collaboration has shown their results in 2015 and 2017 with the LHCb Run\,1 dataset, and thus it was five years passed. 
Then, now, the LHCb collaboration reported their preliminary result of $R_{D}$ and also $R_{D^{\ast}}$ with the LHCb Run\,1 dataset \cite{LHCbRun2},
\begin{align}
    \begin{aligned}
    R_D^{\rm LHCb2022} & =0.441\pm 0.060 \pm 0.066\,,\\
    R_{D^\ast}^{\rm LHCb2022} &= 0.281 \pm 0.018 \pm 0.024\,.
    \end{aligned}
\end{align}
The $\tau$ is reconstructed in $\tau\to\mu\nu\overline\nu$ and the result supersedes the previous result performed in 2015 \cite{Aaij:2015yra}.

\begin{table}[t]
\centering
\newcommand{\bhline}[1]{\noalign{\hrule height #1}}
\renewcommand{\arraystretch}{1.5}
\rowcolors{2}{gray!15}{white}
\addtolength{\tabcolsep}{5pt} 
   \scalebox{1}{
  \begin{tabular}{lccc} 
  \bhline{1 pt}
  \rowcolor{white}
 Experiment  &$R_{D^\ast}$ & $R_{D}$ & Correlation  \\  \hline 
BaBar  \cite{Lees:2012xj,Lees:2013uzd}  & $0.332\pm 0.024\pm 0.018$ & $0.440\pm 0.058 \pm 0.042$ & $-0.27$ \\
Belle \cite{Huschle:2015rga} & $0.293\pm 0.038\pm 0.015$ & $0.375\pm 0.064\pm 0.026$ & $-0.49$\\
Belle \cite{Hirose:2016wfn,Hirose:2017dxl}
& $0.270\pm 0.035^{+0.028}_{-0.025}$ & --& -- \\
Belle \cite{Abdesselam:2019dgh,Belle:2019rba}
& $0.283\pm 0.018\pm 0.014$ & $0.307\pm 0.037\pm 0.016$ & $-0.51$\\
LHCb \cite{Aaij:2017uff,Aaij:2017deq}
& $0.280\pm 0.018\pm 0.029$ & -- & -- \\
LHCb\cite{Aaij:2015yra,LHCbRun2} & $0.281 \pm 0.018 \pm 0.024$ &$0.441\pm 0.060 \pm 0.066$ & $-0.43$ \\ 
\hline
 World average \cite{HFLAV2022fall} & $0.285\pm 0.010\pm 0.008$ &$0.358\pm 0.025\pm 0.012$ & $-0.29$ \\ 
\bhline{1 pt}
   \end{tabular}
}   
\addtolength{\tabcolsep}{-5pt} 
 \caption{ \label{tab:RD_exps}
 Current status of the independent experimental $R_{D^{(\ast)}}$ measurements. 
 The first and second errors are statistical and systematic, respectively. 
 }
\end{table}

In Table~\ref{tab:RD_exps}, we summarize the current status of the $R_{D^{(\ast)}}$ measurements including the new LHCb result. 
It is found that the new LHCb result is consistent with the previous world average evaluated in the HFLAV 2021 report~\cite{HFLAV:2022pwe} 
within the experimental uncertainty.
The combined average of the experimental data gives $p(\chi^2)=32\%$ with $\chi^2/\text{dof}=9.21/8$ for the  $p$-value among all data, 
compared with the previous HFLAV average of $28\%$ with $\chi^2/\text{dof} = 8.8/7$ written in Ref.~\cite{HFLAV:2022pwe}.
The amplified $p$-value indicates 
consistency among the data.
New world averages of the $R_{D^{(\ast)}}$ measurements are \cite{HFLAV2022fall}
\begin{align}
    \begin{aligned}
    R_D &= 0.358\pm 0.025\pm 0.012\,, \\
    R_{D^\ast} & = 0.285\pm 0.010\pm 0.008 \,,
    \end{aligned}
\end{align}
and $R_D$--$R_{D^\ast}$ correlation of $-0.29$.

\begin{figure}[t]
\begin{center}
 \includegraphics[viewport=0 0 800 495, width=36em]{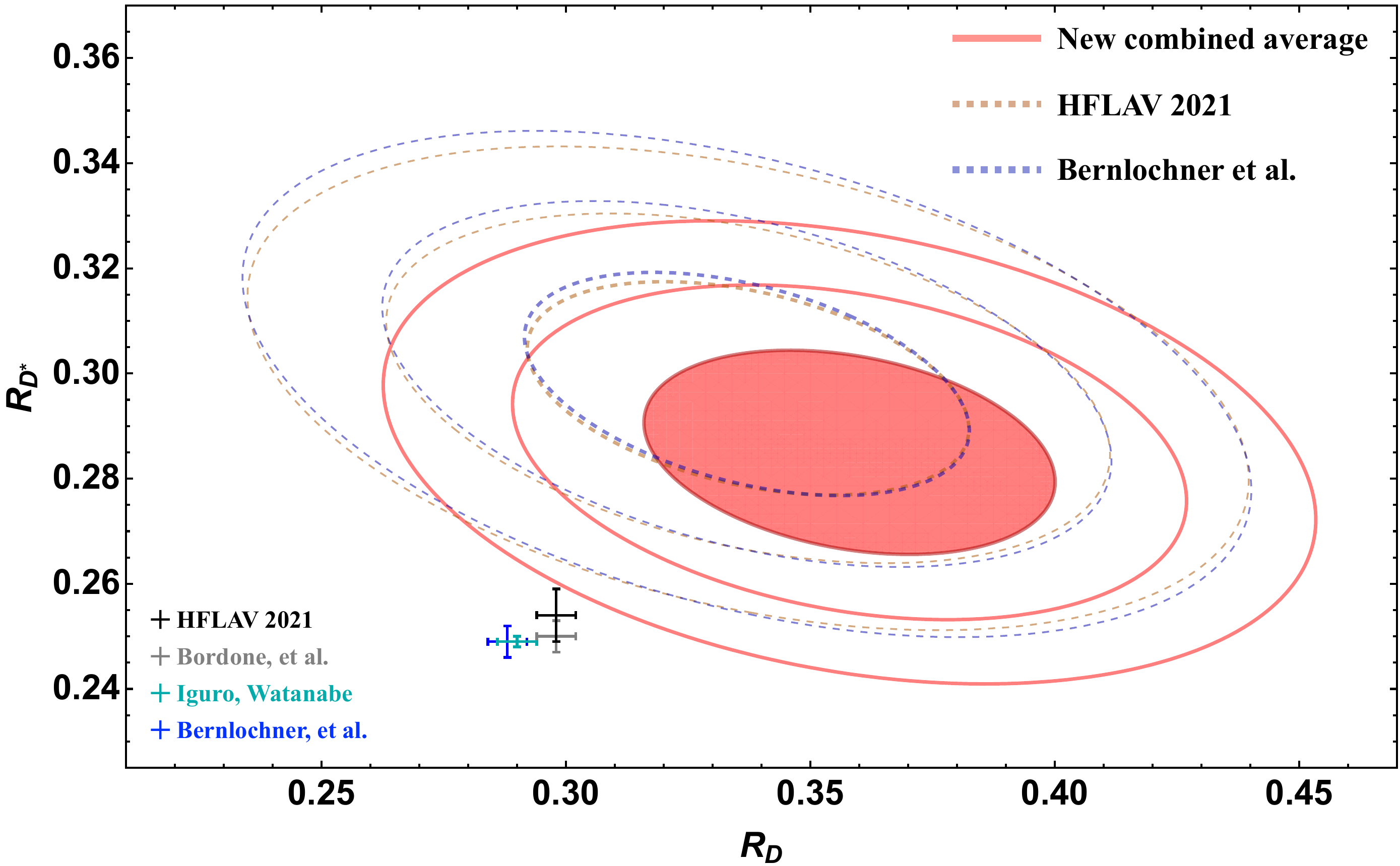}
\end{center}
 \caption{
 \label{fig:RDplot}
 A world average of the latest $R_D$ and $R_{D^*}$ experimental results (red, 1,\,2,\,3$\sigma$ contours), compared with the previous HFLAV 2021 average (dashed orange) \cite{HFLAV:2022pwe} 
 and with Ref.~\cite{Bernlochner:2021vlv} (dashed blue) which includes the nontrivial $D^{\ast\ast}$ contribution. On the other hand, the several SM predictions are shown by crosses \cite{Bernlochner:2022ywh,Iguro:2020cpg, Bordone:2019vic,Bordone:2019guc}.
 }
\end{figure}

Regarding the combined average, an important analysis is given in Ref.~\cite{Bernlochner:2021vlv}. 
The authors pointed out that evaluations of the $D^{\ast\ast}$ distributions in the SM background involve nontrivial correlations that affect the $R_{D^{(\ast)}}$ measurements. 
Their sophisticated study shows that the combined $R_{D^{(\ast)}}$ average is slightly sifted, which is beyond the scope of our work.\footnote{ 
Instead, a comparison among the two previous and new world averages is shown in Fig.~\ref{fig:RDplot}. 
}

\begin{table}[t]
\centering
\newcommand{\bhline}[1]{\noalign{\hrule height #1}}
\renewcommand{\arraystretch}{1.5}
\rowcolors{2}{gray!15}{white}
   \begin{adjustbox}{width=\columnwidth,center}
  \begin{tabular}{lcccccccc} 
  \bhline{1 pt}
  \rowcolor{white}
   Reference & $R_{D}$ & $R_{D^{\ast}}$ & $P_\tau^D$ & $-P_\tau^{D^\ast}$ &
   $F_L^{D^\ast}$ & $R_{\Jpsi}$ & $R_{\Lambda_c}$ & $R_{\Upsilon(3S)}$  \\  \hline 
Bernlochner, {\it et al.} \cite{Bernlochner:2022ywh} & $0.288(4)$ & $0.249(3)$ & -- & -- & -- & -- & -- & -- \\
Iguro, Watanabe \cite{Iguro:2020cpg} & $0.290(3)$ & $0.248(1)$ & $0.331(4)$& $0.497(7)$& $0.464(3)$& -- & -- & -- \\
Bordone, {\it et al.} \cite{Bordone:2019vic,Bordone:2019guc} & $ 0.298(3)$ & $0.250(3)$ & $ 0.321(3)$& $0.492(13)$& $0.467(9)$& --& --& --\\ 
HFLAV2021 \cite{HFLAV:2022pwe} & $0.298 (4)$ & $0.254 (5)$ & --& --& --& --& --& --\\
Refs.~\cite{Harrison:2020nrv,Bernlochner:2018kxh,Aloni:2017eny} & -- & -- & --& --& --& $0.258(4)$& $0.324 (4)$& $0.9948$\\
\hline
Data   & $0.358 (28)$ & $0.285 (13)$ & -- & $0.38 \,^{+0.53}_{-0.55}$ & $0.60 (9)$ &$0.71(25)$ & $0.271 (72)$ & $0.968 (16)$ \\ 
\bhline{1 pt}
   \end{tabular}
\end{adjustbox}
    \caption{ \label{tab:data_SM_summary}
Summary of the SM predictions for the $\Bb \to D^{(\ast)}\tau\overline\nu$ and related observables. 
The current combined results of the experimental measurements are also written in the last line. 
See the main text for the definitions of the observables. 
}
\end{table}
Recent SM predictions for $R_{D^{(\ast)}}^\text{SM}$ have been obtained in Refs.~\cite{Bordone:2019vic,Bordone:2019guc,Iguro:2020cpg,HFLAV:2022pwe} as summarized in Table~\ref{tab:data_SM_summary}. 
The difference of these SM values is mainly due to development of the $\Bb  \to D^{(\ast)}$ form factor evaluations both by theoretical studies and experimental fits, whose details can be found in the literature. 
In our work, we will employ the work of Ref.~\cite{Iguro:2020cpg} as explained soon later. 

A further concern for the SM evaluation is long-distance QED corrections to $\Bb \to D^{(\ast)}\ell\overline\nu$, which remains an open question. 
They depend on the lepton mass as being of $\mathcal{O}[\alpha \ln (m_\ell/m_B)]$ and hence it could provide a few percent correction to violation of the LFU in the semileptonic processes\cite{deBoer:2018ipi,Cali:2019nwp,Isidori:2020eyd,Papucci:2021ztr}.
This will be crucial in future when the Belle~II experiment reaches such an accuracy.

In Fig.~\ref{fig:RDplot}, we show 
the latest average of the $R_D$--$R_{D^\ast}$ along with the several recent SM predictions. 
A general consensus from the figure is that the deviation of the experimental data from the SM expectations still remains.
For instance, applying the SM prediction from \{HFLAV2021~\cite{HFLAV:2022pwe}, Ref.~\cite{Bernlochner:2022ywh}, Ref.~\cite{Iguro:2020cpg}, Refs.~\cite{Bordone:2019vic,Bordone:2019guc}\}, 
one can see $\{3.2 \sigma, 4.0\sigma, 4.1 \sigma, 3.6 \sigma\}$ deviations corresponding to $p\text{-value}=\{1.2 \times  10^{-3}, 6.4 \times 10^{-5}, 4.8\times 10^{-5}, 2.7 \times 10^{-4} \}$ ($\Delta \chi^2 = \{13.8, 19.3, 19.9, 16.4$\} for $2$ degrees of freedom), respectively. 

In addition to these deviations in the LFU measurements, $\tau$- and $D^\ast$-polarization observables in $\Bb \to D^{(\ast)} \tau \overline{\nu}$ also provide us important and nontrivial information.
This is because these observables can potentially help us to pin down the NP structure that causes these deviations~\cite{Tanaka:2010se,Sakaki:2012ft,Duraisamy:2013pia,Duraisamy:2014sna,Becirevic:2016hea,Alok:2016qyh,Ivanov:2017mrj,Colangelo:2018cnj,Bhattacharya:2018kig,Iguro:2018vqb,Blanke:2018yud,Blanke:2019qrx,Becirevic:2019tpx,Hill:2019zja,Alguero:2020ukk,Bhattacharya:2020lfm,Penalva:2021gef,Penalva:2022vxy}.
We refer to the $\tau$ longitudinal-polarization asymmetry in $\Bb \to D^{(\ast)}\tau \overline{\nu}$ and the fraction of the $D^{\ast}$ longitudinal mode in $\Bb \to D^{\ast}\tau \overline{\nu}$ 
as $P_\tau^{D^{(\ast)}}$ and $F_{L}^{D^{\ast}}$, respectively. 
See Refs.~\cite{Tanaka:2012nw,Asadi:2018sym,Iguro:2018vqb} for their explicit definitions. 

In recent years, the first measurements for some of the above polarization observables have been reported by the Belle collaboration.  
It is summarized as $P_\tau^{D^\ast} = -0.38 \pm 0.51 \,^{+0.21}_{-0.16}$~\cite{Hirose:2016wfn} and $F_L^{D^\ast} = 0.60 \pm 0.08 \pm 0.04$~\cite{Abdesselam:2019wbt}. 
See also Table~\ref{tab:data_SM_summary}. 
Although the experimental uncertainty in $F_L^{D^\ast}$ is still large, 
this result already has important implications for a tensor-operator NP as pointed out in Ref.~\cite{Iguro:2018vqb}. 
Although $P_\tau^{D}$ is the most striking observable to disentangle the leptoquark (LQ) scenarios that can explain the discrepancy, the experimental study does not exist so far.

Note that the $D^{\ast}$ longitudinal polarization in the electron mode has also been measured, 
$F_L^{D^\ast}(\Bb{}^0 \to D^{\ast +}e \overline{\nu}) \equiv F_L^{D^\ast\!, e} = 0.56 \pm  0.02$~\cite{Abdesselam:2019wbt}.
This is comparable to the SM prediction of $0.534 \pm 0.002$ \cite{Iguro:2020cpg}. 
The $F_L^{D^\ast}$ and $F_L^{D^\ast\!, e}$ measurements have the same level of significance (1.5$\sigma$ and 1.3$\sigma$, respectively).

\subsection{Preliminaries of our analysis}

Main points of this paper are that 
(i) we provide state-of-the-art numerical formulae for the observables relevant to the semi-tauonic $B$ decays and 
(ii) we revisit to perform global fits to the available $R_{D^{(\ast)}}$ measurements with respect to NP interpretations. 
It will be given by incorporating following updates and concerns:
\begin{itemize}
 \item 
 The preliminary result from the LHCb collaboration is encoded in our world average as shown in Table~\ref{tab:RD_exps}. 
 \item 
 The recent development of the $\Bb\to D^{(\ast)}$ transition form factors is taken into account.
 It is described by the heavy quark effective theory (HQET) taking higher-order corrections up to $\mathcal{O}(\Lambda_{\rm QCD}^2/m_c^2)$ as introduced in Refs.~\cite{Jung:2018lfu,Bordone:2019vic}. 
 We follow the result from the comprehensive theory+experiment fit analysis as obtained in Ref.~\cite{Iguro:2020cpg}.\footnote{
 To be precise, we employ the ``(2/1/0) fit'' result, preferred by their fit analysis.
 See the reference for details.} 
 \item
 The recent study of Ref.~\cite{Bernlochner:2022ywh} introduced an approximation method to reduce independent parameters involving the $\mathcal{O}(\Lambda_{\rm QCD}^2/m_c^2)$ corrections in HQET. 
 Although this affects some of the parameter fits for the form factors, 
 readers can find in Table~\ref{tab:data_SM_summary} that our reference values of Ref.~\cite{Iguro:2020cpg} are consistent with those of Ref.~\cite{Bernlochner:2022ywh} as for $R_{D^{(\ast)}}$. 
 Hence we do not take this approximation in our work. 
 \item
 Recently, Fermilab--MILC Collaborations~\cite{Bazavov:2021bax} presented the first lattice result of the form factors for $\Bb \to D^\ast \ell \overline \nu$ at nonzero recoil, 
 with which one obtains $R_{D^\ast}^{\rm SM}  = 0.2484 \pm 0.0013$ and again readers can see the consistency with our reference. 
 Since this preliminary result needs to be finalized and to be compared with upcoming lattice results from the other collaborations such as JLQCD and HPQCD, 
 we do not include this update in our analysis. 
 \item
 Another way of constraining the form factor has been discussed in Refs.~\cite{Martinelli:2021frl,Martinelli:2021onb,Martinelli:2021myh}. 
 It is free from the parameterization method and obtained by a general property of the unitary bound. 
 They found $R_D^{\rm SM} =0.296\pm0.008$ and $R_{D^\ast}^{\rm SM}=0.261\pm0.020$ ($R_{D^\ast}^{\rm SM}=0.275\pm0.021$ by taking the Fermilab--MILC result~\cite{Bazavov:2021bax}), 
 slightly larger, but still consistent within the uncertainty. 
 We do not consider this case. 
 \item
 Indirect LHC bounds from the high-$p_T$ mono-$\tau$ searches with large missing transverse energy~\cite{Greljo:2018tzh,Dumont:2016xpj,Altmannshofer:2017poe,Iguro:2017ysu,Abdullah:2018ets,Iguro:2018fni,Baker:2019sli,Marzocca:2020ueu,Iguro:2020keo,Endo:2021lhi,Jaffredo:2021ymt} 
 are concerned.  
 We impose the result of Ref.~\cite{Iguro:2020keo} that directly constrains the NP contributions to the $b\to c\tau\overline\nu$ current and accounts for the NP-scale dependence on the LHC bound, 
 which is not available by the effective-field-theory description.
 Requiring an additional $b$-tagged jet also helps to improve the sensitivity~\cite{Marzocca:2020ueu,Endo:2021lhi}.
 We will see how it affects the constraints in the leptoquark scenarios.
 \item
 Similar sensitivity can be obtained by the $\tau\tau$ final state~\cite{ATLAS:2020zms,CMS:2022goy}.
 It is noted that the about three standard deviation is reported by the CMS collaboration~\cite{CMS:2022goy}, which would imply the existence of leptoquark,
 while the ATLAS result~\cite{ATLAS:2020zms} has not found the similar excess.
 We need the larger statistics to confirm it, and thus we do not include the constraint to be conservative.
\end{itemize}
In addition to the above points, we also investigate the following processes that are directly/indirectly related to the $b \to c\tau\overline\nu$ current:
\begin{itemize}
 \item 
 The LFU in $B_c\to \Jpsi\, l\,\overline\nu$ decays is connected to $R_{D^{(*)}}$. 
 The LHCb collaboration has measured the ratio 
 $R_{\Jpsi}\equiv \mathcal{B}(B_c\to \Jpsi\,\tau\,\overline\nu)/\mathcal{B}(B_c\to \Jpsi\,\mu\,\overline\nu) = 0.71\pm 0.17\pm 0.18$~\cite{Aaij:2017tyk}.
 Although the current data includes such a large uncertainty,
 it would be useful in future to test some NP scenarios for the sake of the $R_{D^{(*)}}$ anomalies. 
 We update the numerical formula for $R_{\Jpsi}$ in the presence of general NP contributions and put a prediction from our fit study. 
 \item 
 The $\Upsilon$ leptonic decays, $\Upsilon\to l^+ l^-$,  are potentially connected to $R_{D^{(\ast)}}$ once one specifies NP interactions to the bottom quark and leptons. 
 Although the SM contribution comes from a photon exchange, it is suppressed by the $\Upsilon$ mass squared.
 The sensitivity to NP is, therefore, not completely negligible,
 and the LFU of $R_{\Upsilon(nS)}\equiv  \mathcal{B}(\Upsilon(nS) \to \tau^+ \tau^-) / \mathcal{B}(\Upsilon(nS) \to \ell^+ \ell^-)$ 
 can be an important cross check of the $R_{D^{(*)}}$ anomalies.
 Furthermore, 
 the BaBar collaboration has reported 
 a result which slightly violates the LFU: $R_{\Upsilon(3S)}= 0.966 \pm 0.008 \pm 0.014$~\cite{Lees:2020kom}. 
 We investigate the theoretical correlations in several NP models.
\end{itemize}

This paper is organized as follows. 
In Sec.~\ref{sec_GF}, we put the numerical formulae for the relevant observables in terms of the effective Hamiltonian.
We also summarize the case for single operator analysis. 
In Sec.~\ref{sec:Fit}, based on the generic study with renormalization-group running effects, 
we obtain relations among $R_{D}$, $R_{D^{\ast}}$, and $F_{L}^{D^{\ast}}$ in the LQ models and discuss their potential to explain the present data. 
Relations to the $\tau$ polarizations are also discussed.
In Sec.~\ref{sec:Upsi_decay}, we also investigate the LFU violation in the $\Upsilon$ decays and show its correlation with $b\to c \tau \overline \nu$ observables.
Finally, we conclude in Sec.~\ref{sec:conclusion}.

\section{General formulae for the observables}
\label{sec_GF}

At first, we describe general NP contributions to $b \to c\tau\overline\nu$ in terms of the effective Hamiltonian.
The operators relevant to the processes of interest are described as\footnote{The different naming scheme of the operators are often used \cite{Sakaki:2013bfa,Huang:2018nnq,Kou:2018nap}.
Our $C_{V_L},\,C_{V_R},\,C_{S_L}$, and $C_{S_R}$ correspond to $C_{V_1},\,C_{V_2},\,C_{S_2}$, and $C_{S_1}$, respectively.}
\begin{align}
 \label{eq:Hamiltonian}
 {\cal {H}}_{\rm{eff}}= 2 \sqrt2 G_FV_{cb}\biggl[ (1+C_{V_L})O_{V_L}+C_{V_R}O_{V_R}+C_{S_L}O_{S_L}+C_{S_
R}O_{S_R}+C_{T}O_{T}\biggl]\,,
\end{align}
with
\begin{align}
 &O_{V_L} = (\overline{c} \gamma^\mu P_Lb)(\overline{\tau} \gamma_\mu P_L \nu_{\tau})\,,& 
 &O_{V_R} = (\overline{c} \gamma^\mu P_Rb)(\overline{\tau} \gamma_\mu P_L \nu_{\tau})\,,&\nonumber \\
 &O_{S_L} = (\overline{c}  P_Lb)(\overline{\tau} P_L \nu_{\tau})\,,&
 &O_{S_R} = (\overline{c}  P_Rb)(\overline{\tau} P_L \nu_{\tau})\,,& \label{eq:operator} \\
 &O_{T} = (\overline{c}  \sigma^{\mu\nu}P_Lb)(\overline{\tau} \sigma_{\mu\nu} P_L \nu_{\tau}) \,,&
 \nonumber
\end{align}
where $P_L=(1-\gamma_5)/2$ and $P_R=(1+\gamma_5)/2$. 
The NP contribution is encoded in the Wilson coefficients (WCs) of $C_X$, normalized by the SM factor of $2 \sqrt2 G_FV_{cb}$. 
The SM corresponds to 
$C_{X} = 0$ for $X=V_{L,R}$, $S_{L,R}$, and $T$ 
in this description. 
We assume that the light neutrino is always left-handed,
and NP contributions are relevant to only the  third-generation neutrino ($\nu_\tau$), for simplicity.\footnote{%
See Refs. \cite{Iguro:2018qzf,Asadi:2018wea,Greljo:2018ogz,Robinson:2018gza,Babu:2018vrl,Mandal:2020htr,Penalva:2021wye} for models with the right-handed neutrino $\nu_R$.
It is noted that the $W^\prime$ is necessarily accompanied by $Z^\prime$ and thus the recent di-$\tau$ resonance search \cite{ATLAS:2020zms,CMS:2022goy} excludes the $W_R^\prime$ explanation \cite{IguroKEK}.}

Note that the leading $SU(2)_L \times U(1)_Y$ invariant operator, to generate the LFU violated type of the $O_{V_2}$ form, is given in dimension-eight 
as $(\overline{c}_R \gamma^\mu b_R)(\overline{L}^3 \gamma_\mu \tau^A L^3)(\tilde H \tau^A H)$. 
This implies that $C_{V_2}$ in a NP model necessarily has an additional suppression
compared with the other operators generated from dimension-six operators. 
See Ref.~\cite{Asadi:2019zja} for a NP model that can generate the  $C_{V_2}$ contributions to $R_{D^{(\ast)}}$.

In the following parts, the observables for $\Bb \to D^{(\ast)} \tau \overline{\nu}$, $B_c \to \tau \overline{\nu}$, and $B_c\to \Jpsi\, \tau\,\overline{\nu}$ 
are evaluated with Eq.~\eqref{eq:Hamiltonian} at the scale $\mu = \mu_b = 4.18\,\text{GeV}$. 
The process $\Upsilon(nS) \to l^+ l^-$ will be described  in detail in Sec.~\ref{sec:Upsi_decay}.

\subsection*{\underline{$\Bb \to D^{(\ast)} \tau \overline{\nu}$}}
In this work, we follow analytic forms of the differential decay rates for $\Bb \to D^{(\ast)} \tau \overline{\nu}$ obtained in Refs.~\cite{Sakaki:2013bfa,Sakaki:2014sea}. 
Regarding the form factors, we employ the general HQET based description~\cite{Bordone:2019vic}, 
in which the heavy quark expansions~\cite{Caprini:1997mu,Bernlochner:2017jka} are taken up to 
NLO for $\epsilon_a = \alpha_s/\pi$, $\epsilon_b = \overline\Lambda/(2m_b)$ and NNLO for $\epsilon_c = \overline\Lambda/(2m_c)$ by recalling the fact $\epsilon_a \sim \epsilon_b \sim \epsilon_c^2$. 
Thanks to HQET property, the form factors for the different Lorenz structures of the NP operators are connected to that for the SM current, 
which enables us to evaluate the NP contributions to the observables.

Two parametrization models have been considered with respect to the $z = (\sqrt{w+1} -\sqrt 2) / (\sqrt{w+1} +\sqrt 2)$ expansions for the form factors in this description, 
with which the most general fit analyses of the form-factor parameters and $|V_{cb}|$ have been performed in Ref.~\cite{Iguro:2020cpg}. 
For the present work, we take the $(2/1/0)$ model with a minor update and apply the updated fit result based on Ref.~\cite{Iguro:2020cpg}.

We have evaluated the ratio observables, $R_{D^{(\ast)}}$, $P_\tau^{D^{(\ast)}}$ and $F_{L}^{D^{\ast}}$, 
for the case of the effective Hamiltonian of Eq.~\eqref{eq:Hamiltonian} at the scale $\mu = \mu_b$. 
In the end, we find the following updated numerical formulae,
\begin{align}
 \label{eq:RD}
 \frac{R_D}{R_{D}^\textrm{SM}} =
 & ~|1+C_{V_L}+C_{V_R}|^2  + 1.01|C_{S_L}+C_{S_R}|^2 + 0.84|C_{T}|^2  \nonumber \\[-0.5em]
 & + 1.49\textrm{Re}[(1+C_{V_L}+C_{V_R})(C_{S_L}^*+C_{S_R}^*)]  + 1.08\textrm{Re}[(1+C_{V_L}+C_{V_R})C_{T}^*] \,, 
 \\[1em]
 %
 \label{eq:RDs}
 \frac{ R_{D^{\ast}}}{R_{D^\ast}^\textrm{SM}} =
 & ~|1+C_{V_L}|^2 + |C_{V_R}|^2  + 0.04|C_{S_L}-C_{S_R}|^2 + 16.0|C_{T}|^2 \nonumber \\[-0.5em]
 & -1.83\textrm{Re}[(1+C_{V_L})C_{V_R}^*]  - 0.11\textrm{Re}[(1+C_{V_L}-C_{V_R})(C_{S_L}^*-C_{S_R}^*)] \nonumber \\
 & -5.17\textrm{Re}[(1+C_{V_L})C_{T}^*] + 6.60\textrm{Re}[C_{V_R}C_{T}^*] \,, 
 \\[1em]
%
 \label{eq:PtauD}
 \frac{P_\tau^{D}}{P_{\tau,\,\textrm{SM}}^{D}} = 
 & \left(\frac{R_D} {R_D^\text{SM}}\right)^{-1} \!\!\!\times \Big( |1+C_{V_L}+C_{V_R}|^2  + 3.04|C_{S_L}+C_{S_R}|^2 + 0.17|C_{T}|^2 \nonumber \\
 & + 4.50\textrm{Re}[(1+C_{V_L}+C_{V_R})(C_{S_L}^*+C_{S_R}^*)]  -1.09\textrm{Re}[(1+C_{V_L}+C_{V_R})C_{T}^*] \Big)  \,, 
 \\[1em]
 \label{eq:PtauDs}
 \frac{P_\tau^{D^{\ast}}} {P_{\tau,\,\textrm{SM}}^{D^{\ast}}} = 
 & \left(\frac{R_{D^{\ast}}}  {R_{D^{\ast}}^\text{SM}}\right)^{-1} \!\!\!\times \Big( |1+C_{V_1}|^2  + |C_{V_2}|^2  - 0.07|C_{S_1}- C_{S_2}|^2 - 1.85 |C_{T}|^2 \nonumber \\
 & 
 - 1.79 \textrm{Re}[(1+C_{V_L})C_{V_R}^{\ast}]  
 + 0.23 \textrm{Re}[(1+C_{V_L}- C_{V_R})(C_{S_L}^* -C_{S_R}^*)]  \nonumber \\[0.5em]
 & 
 - 3.47 \textrm{Re}[(1+C_{V_L}) C_{T}^*]
 + 4.41  \textrm{Re}[C_{V_R} C_{T}^*] \Big)\,, 
 \\[1em]
%
 \label{eq:FLDs}
 \frac{F_L^{D^*}}{F_{L,\,\textrm{SM}}^{D^{\ast}}} = 
 & \left(\frac{R_{D^*}}{R_{D^*}^\text{SM}}\right)^{-1} \!\!\!\times \Big( |1+C_{V_L}-C_{V_R}|^2  + 0.08|C_{S_L}-C_{S_R}|^2 + 6.90|C_{T}|^2 \nonumber \\
 & - 0.25\textrm{Re}[(1+C_{V_L}-C_{V_R})(C_{S_L}^*-C_{S_R}^*)]  -4.30\textrm{Re}[(1+C_{V_L}-C_{V_R})C_{T}^*] \Big) \,,
\end{align}
which can be compared with those in the  literature~\cite{Feruglio:2018fxo,Asadi:2018wea,Blanke:2018yud,Iguro:2018vqb,Mandal:2020htr}. 
The SM predictions are obtained as\footnote{
We updated the fit analysis with the modification of the formula for unitarity bound~\cite{Caprini:1997mu}, pointed out by Ref.~\cite{Bigi:2017jbd}. 
It only affects the last digits of the SM predictions, though.
} 
\begin{align}
    \begin{aligned}
   R_{D}^\textrm{SM} &= 0.290 \pm 0.003\,,\\[0.5em]
   R_{D^\ast}^\textrm{SM}& = 0.248\pm 0.001\,,\\[0.5em]
   P_{\tau,\,\textrm{SM}}^{D} &= 0.331 \pm 0.004\,,\\[0.5em]
   P_{\tau,\,\textrm{SM}}^{D^{\ast}} &= -0.497 \pm 0.007\,,\\[0.5em]
   F_{L,\,\textrm{SM}}^{D^{\ast}} &= 0.464 \pm 0.003\,.
    \end{aligned}
\end{align}

Furthermore, we have checked uncertainties of the above numerical coefficients in the formulae,
based on the fit result from Ref.~\cite{Iguro:2020cpg}. 
The tensor (scalar) terms involve $\sim4\%\, (10\%)$ uncertainties for the $D$ ($D^*$) mode,  while the others contain less than $1\%$ errors. 
At present, they are not significant and thus neglected in our following study.

\subsection*{\underline{$B_c \to \tau \overline{\nu}$}}
The significant constraint 
on the scalar operators $O_{S_{L,R}}$ 
comes from the $B_c$ lifetime measurements ($\tau_{B_c}$) \cite{Beneke:1996xe,Alonso:2016oyd,Celis:2016azn,Watanabe:2017mip,Aebischer:2021ilm}:
the branching ratio of $B_c^- \to \tau \overline\nu$, which has not yet been observed, 
is significantly amplified by the NP scalar interactions, and the branching ratio is constrained from measured $\tau_{B_c}$ \cite{Zyla:2020zbs}.
We obtain an upper bound on the WCs as
\begin{align}
 \label{eq:Bc}
 \left| 1 + C_{V_L} - C_{V_R} - 4.35\, (C_{S_L} - C_{S_R}) \right|^2 = \frac{\mathcal{B}(B_c\to \tau\overline\nu)}{\mathcal{B}(B_c\to \tau\overline\nu)_{\rm SM}} < 
27.1\left( \frac{\mathcal{B}(B_c\to \tau\overline\nu)_{\rm UB}}{0.6}\right)\,,
\end{align}
with $\mathcal{B}(B_c\to \tau\overline\nu)_{\rm SM} \simeq 0.022$.
Here, $|V_{cb}| = \left(41.0\pm 1.4\right) \times 10^{-3}$ is used \cite{Zyla:2020zbs}. 
The $b$ and $c$ quark mass inputs, which are relevant for scalar contributions, are taken as $m_b(\mu_b)=(4.18 \pm 0.03)\,\text{GeV}$ and $m_c(\mu_b)=(0.92\pm0.02)\,\text{GeV}$. 
Reference \cite{Alonso:2016oyd} evaluated that 
the upper bound (UB) on the branching ratio from $\tau_{B_c}$ is 
$\mathcal{B}(B_c\to \tau\overline\nu)_{\rm UB}=0.3$.
However, it is pointed out by Ref.~\cite{Blanke:2018yud} and later confirmed by Ref.~\cite{Aebischer:2021ilm} 
that there is a sizeable charm-mass dependence on the $B_c$ decay rate because the dominant contribution comes from the charm-quark decay into strange within the $B_c$ meson.
A conservative bound is set by Ref.~\cite{Blanke:2018yud}
as $\mathcal{B}(B_c\to \tau\overline\nu)_{\rm UB}=0.6$.

One should note that 
more aggressive bound $\mathcal{B}(B_c \to \tau \overline{\nu})_{\rm UB} = 0.1$ has been obtained in Ref.~\cite{Akeroyd:2017mhr} by using LEP data.
However, it is pointed out that $p_T$ dependence 
of the fragmentation function, $b \to B_c$, has been entirely overlooked, 
and thus the bound must be overestimated by several factors \cite{Blanke:2018yud,Bardhan:2019ljo,Blanke:2019qrx}. 
Although the CEPC and FCC-ee experiments
are in planning stages,
the future Tera-$Z$ machines can directly measure $\mathcal{B}(B_c\to\tau\overline\nu)$ 
at O(1$\%$) level \cite{Zheng:2021xuq,Amhis:2021cfy}.

Thanks to the conservative bound, the left-handed scalar operator, $C_{S_L}$ comes back to the game.
For instance a general two-Higgs doublet model is a viable candidate and readers are referred to Refs.\cite{Iguro:2022uzz,Blanke:2022pjy}.

\subsection*{\underline{$B_c\to \Jpsi\, \tau\,\overline{\nu}$}}
We follow the form factor description from the recent lattice result of Ref.~\cite{Harrison:2020gvo} for $B_c\to \Jpsi\, \tau\,\overline{\nu}$. 
We also take $m_b(\mu_b)$ and $m_c(\mu_b)$ for the scalar and tensor sectors as aforementioned. 
The formula for $R_{\Jpsi}$ is given as 
\begin{align}
\begin{aligned}
 \frac{ R_{\Jpsi}}{R_{\Jpsi}^\textrm{SM}} =
 & ~|1+C_{V_L}|^2 + |C_{V_R}|^2  +0.04|C_{S_L}-C_{S_R}|^2 + 14.7|C_{T}|^2  \\
 & -1.82\textrm{Re}[(1+C_{V_L})C_{V_R}^*]  - 0.10\textrm{Re}[(1+C_{V_L}-C_{V_R})(C_{S_L}^*-C_{S_R}^*)]  \\[0.5em] 
 & -5.39\textrm{Re}[(1+C_{V_L})C_{T}^*] + 6.57\textrm{Re}[C_{V_R}C_{T}^*] \,, 
 \end{aligned}
\end{align}
where we take $R_{\Jpsi}^\textrm{SM} = 0.258 \pm 0.004$~\cite{Harrison:2020nrv}. 
The coefficients potentially have $10$--$20\%$ uncertainties for $C_{S_{L,R}}$ and $C_T$, while a few percent for $C_{V_{L,R}}$.

\section{Fit analysis}
\label{sec:Fit}
In this paper, several NP scenarios are investigated in accordance with the following steps: 
\begin{enumerate}
 \item 
 The measurements of $R_{D}$,  $R_{D^{*}}$  and $F_L^{D^*}$ are taken in the $\chi^2$ fit, 
 and then the favored regions for the NP parameter space are obtained. 
 \item 
 We then check whether the above solutions are consistent with the other relevant observables, such as the $B_c$ lifetime and the LHC bound. 
 \item
 Furthermore, we evaluate NP predictions on $P_\tau^{D}$, $P_\tau^{D^{*}}$ and $R_{\Jpsi}$.
 \item
 If applicable, a combined study with $R_{\Upsilon(3S)}$ is discussed. 
\end{enumerate}
The $\chi^2$ fit function is defined as 
\begin{align}
    \chi^2\equiv\sum_{i,j} (O^{\rm{theory}} 
-O^{\rm{exp}})_i ~{\rm Cov}^{-1}_{ij} 
~(O^{\rm{theory}}-O^{\rm exp})_{j} \,, 
\end{align}
where we take into account the $R_{D^{(*)}}$ and $F_L^{D^*}$ measurements for $O^{\rm{exp}}$ reported by BaBar, LHCb, and Belle collaborations. 
The covariance is given as ${\rm Cov}_{ij} = \Delta O^{\rm{exp}}_i \rho_{ij} \Delta O^{\rm{exp}}_j + \Delta O^{\rm{theory}}_i \delta_{ij} \Delta O^{\rm{theory}}_j$, 
where correlation $\rho_{ij}$ is given as in Table~\ref{tab:RD_exps} while $\rho_{ij} = \delta_{ij}$ among the independent measurements. 
For every observables, we have the theory formulae $O^{\rm{theory}}$ as shown in Sec.~\ref{sec_GF}, 
and hence obtain best fit values in terms of the WCs as defined in Eq.~\eqref{eq:Hamiltonian}.

Given the SM predictions as $R_{D}^\textrm{SM} = 0.290 \pm 0.003$, $R_{D^\ast}^\textrm{SM} = 0.248 \pm 0.001$, and $F_{L,\,\textrm{SM}}^{D^{\ast}} = 0.464 \pm 0.003$, 
we obtain $\chi^2_\text{SM} = 22.4$ (corresponding to $4.0\sigma$) implying a large deviation from the SM. 
Recall that this chi-square contains the $F_L^{D^*}$ fit which enlarges the value compared with the $R_{D^{(*)}}$ fit shown in Sec.~\ref{sec:intro}.
In order to see how NP scenarios improve the fit, 
we use ``Pull'' value (defined in, \eg, Refs.~\cite{Descotes-Genon:2015uva,Blanke:2018yud}). 
For cases of the single WC fits, the Pull is equivalent to
\begin{align}
 \text{Pull} \equiv  \sqrt{\chi^2_\text{SM} - \chi^2_\text{NP-best}} ~(\sigma) \,, 
\end{align}
such that we can discuss quantitative comparisons among the NP scenarios.

\begin{table}[t]
\centering
\newcommand{\bhline}[1]{\noalign{\hrule height #1}}
\renewcommand{\arraystretch}{1.5}
\rowcolors{2}{gray!15}{white}
\addtolength{\tabcolsep}{5pt} 
 \begin{adjustbox}{width=\columnwidth,center}
  \begin{tabular}{cccccc} 
  \bhline{1 pt}
  \rowcolor{white}
  &$|C_{V_L}|$ & $|C_{V_R}|$ & $|C_{S_L}|$ & $|C_{S_R}|$ & $|C_{T}|$  \\    
  \hline
 EFT ($>10\,\text{TeV}$) & $  0.32~(0.09) $ & $  0.33~(0.09) $ &$   0.55~(0.14) $ & $ 0.55~(0.15) $ &  $  0.17~(0.04)$ \\ 
LQ ($4\,\text{TeV}$) & $ 0.36~(0.10)  $ & $0.40~(0.10)$ & $0.74~(0.17)$ & $0.67~(0.18)$  & $0.22~(0.05)$  \\ 
LQ ($2\,\text{TeV}$) & $  0.42~(0.12) $ & $ 0.51~(0.15) $ & $ 0.80~(0.22) $ & $ 0.77~(0.22) $ & $ 0.30~(0.07)$ \\ 
\bhline{1 pt}
 \end{tabular}
 \end{adjustbox}
\addtolength{\tabcolsep}{-5pt} 
    \caption{\label{tab:LHC_bound_one}
The $95\%$ CL upper bounds 
on the WCs at the $\mu = \mu_b$ scale from the LHC analysis of the $\tau + \text{missing}$ search \cite{Iguro:2020keo}. 
The future prospects with $b$-tagged jet + $\tau \nu$ final state assuming 3\,ab$^{-1}$ of accumulated data are given in the parenthesis \cite{Endo:2021lhi}.
The NP mass scale is shown as $M_{\rm LQ} = 2$\,TeV, $4\,$TeV and $\Lambda_{\rm EFT} >10\,$TeV.
}
\end{table}
Regarding the LHC bound to be compared with the above fit result, we refer to the result from Ref.~\cite{Iguro:2020keo}, in which the $\tau + \text{missing}$ searches have been analyzed. 
Their result is shown in Table~\ref{tab:LHC_bound_one}, where we give the $95\%$ CL upper limit at the $\mu_b$ scale.\footnote{
Note that Table~2 of Ref.~\cite{Iguro:2020keo} shows the
LHC bound at $\mu = \Lambda_\text{LHC}$. 
}
It should be emphasized that the LHC bound on the WC has a non-negligible mediator mass dependence, see Ref.~\cite{Iguro:2020keo} for details. 
This feature is indeed crucial for some NP scenarios as will be seen later.
Furthermore it is pointed out that the charge asymmetry of the $\tau$ lepton will improve the bound on $C_X$.

\subsection{EFT: single operator scenario}
\label{sec:singleWCfit}
We begin with the single NP operator scenarios based on the effective field theory (EFT) of Eq.~\eqref{eq:Hamiltonian}.
Assuming the WC to be real, we immediately obtain the fit results with the Pull values
and predictions of $P_\tau^D,~P_{\tau}^{D^\ast}$ and $R_{\Jpsi}$ as shown in Table~\ref{tab:fit:singlereal}. 
The allowed regions from the $B_c$ lifetime and current LHC bounds are listed as well.

For all the NP scenarios, we can see much improvement on the fit compared with the SM.
A significant change from the previous conclusion (before the new LHCb result~\cite{LHCbRun2} came up~\cite{Blanke:2019qrx,Murgui:2019czp,Shi:2019gxi}) is that 
the $C_{S_R}$ scenario {\em becomes consistent with the data within 95\% CL}, \ie, $\chi^2_{best} < 8.0$ (for three observed data).
Furthermore, the $C_{S_R} (\mu_b) \simeq 0.2$ solution indicates the second best Pull. 
Unfortunately, it is known that the usual type~II  two-Higgs doublet model (2HDM) cannot achieve this solution because
the sign of $C_{S_R}$ must be negative: $C_{S_R} = - m_bm_\tau \tan^2\beta/m_{H^\pm}^2 <0$.
It is noted that even in the generic 2HDM, sizable $C_{S_R}$ contribution is difficult due to constraints from $\Delta M_{s}$ and the LHC search~\cite{Iguro:2017ysu,Faroughy:2016osc}.
Instead, the $C_{S_L}$ scenario is likely to explain the present data.
This is the same feature with the previous fit result before the LHCb data is included. 
The $C_{V_L}$ scenario well explains the present data,  while $C_{V_R}$ gives a lower Pull. 
The $C_T$ solution gives unique predictions on the other observables, which may be able to identify the NP scenario, and
it predicts a large shift of $F_L^{D^\ast}$ with opposite direction from the present measurement~\cite{Abdesselam:2019wbt,Iguro:2018vqb}.

\begin{table}[t]
\centering
\newcommand{\bhline}[1]{\noalign{\hrule height #1}}
\renewcommand{\arraystretch}{1.5}
\rowcolors{3}{white}{gray!15}
   \begin{adjustbox}{width=\columnwidth,center}
  \begin{tabular}{l c c|c c|c c c} 
  \bhline{1 pt}
 \multirow{2}{*}{ } & 
 \multirow{2}{*}{Pull~[$\chi^2_\text{best}$]} & 
 \multirow{2}{*}{Fitted $C_X$} &
 \multicolumn{2}{c|}{Allowed region of $C_X$}&
 \multicolumn{3}{c}{Predictions $(\Delta\chi^2 \le 1)$} \\
 %
 & & 
 & $B_c \to \tau \overline \nu$ & LHC & $P_\tau^D$ & $-P_\tau^{D^*}$ & $R_{\Jpsi}$  \\
 \hline
 SM &
 --${~[22.4]}$ &
 -- &
 -- &  -- &
 $0.331 \pm 0.004$ &
 $0.497 \pm 0.007$ &
 $0.258 \pm 0.004$ \\
 \hline
 $C_{V_L}$ &
 ${4.4~[2.8]}$ &
 ${+0.08(2)}$ &
 very loose &  $[-0.32,0.32]$ &
 ${[0.331, 0.331]}$ &
 ${[0.497, 0.497]}$ &
 ${[0.293, 0.313]}$ \\
 %
 %
 $C_{V_R}$ &
 ${1.9~[18.8]}$ &
 ${-0.05(3)}$ &
 very loose &  $[-0.33,0.33]$ &
 ${[0.331, 0.331]}$ &
 ${[0.496, 0.497]}$ &
 ${[0.271, 0.299]}$ \\
 %
 %
 $C_{S_L}$ &
 ${3.0~[13.3]}$ &
 ${+0.17(5)}$ &
 $[-0.94,1.4]$ &  $[-0.55,0.55]$ &
 ${[0.440, 0.518]}$ &
 ${[0.517, 0.533]}$ &
 ${[0.254, 0.256]}$ \\
 %
 %
 $C_{S_R}$ &
 ${3.8~[7.9]}$ &
 ${+0.20(5)}$ &
 $[-1.4,0.94]$ &  $[-0.55,0.55]$ &
 ${[0.463, 0.529]}$ &
 ${[0.454, 0.471]}$ &
 ${[0.261, 0.263]}$ \\
 %
 %
 $C_T$ &
 ${3.4~[10.6]}$ &
 ${-0.03(1)}$ &
 -- &  $[-0.17,0.17]$ &
 ${[0.347, 0.360]}$ &
 ${[0.459, 0.476]}$ &
 ${[0.286, 0.309]}$ \\
\bhline{1 pt}
 \end{tabular}
 \end{adjustbox}
 \caption{
 The fit results of the single NP operator scenarios assuming real WCs.
 The WCs are given at the $\mu_b$ scale. 
 The allowed ranges of WC from the $B_c$ lifetime and the current LHC bounds are also shown (``very loose'' represents very weak bounds).
 Fitted WCs and predictions of $P_\tau^D,~P_{\tau}^{D^\ast}$ and $R_{\Jpsi}$
 are in the range of $\Delta\chi^2 = \chi^2 - \chi^2_\text{best} \leq 1$. 
 }
 \label{tab:fit:singlereal}
\end{table}  

Once we allow complex values of WCs,
the complex $C_{V_R}$, $C_{S_L}$, and $C_T$ scenarios improve the fits such as 
\begin{align}
 & C_{V_R} \simeq 0.02 \pm i\, 0.43 & & \text{Pull} = 4.1 \,,& \label{eq:complex_start} \\
 & C_{S_L} \simeq -0.88 \pm i\, 0.88 & & \text{Pull} = 4.3 \,,& \label{eq:fit_CSL_complex}\\
 & C_{T} \simeq 0.06 \pm i\, 0.16 & & \text{Pull} = 3.3 \,,\label{eq:complex_end}& 
\end{align}
while the complex $C_{V_L}$ and $C_{S_R}$ scenarios give the same Pulls, compared with those with the real WC scenarios. 
The complex $C_T$ scenario has a similar Pull with the real $C_T$ case. 
The complex $C_{V_R}$ result at the above best fit point is, however, not consistent with the LHC bound for the case of EFT, $|C_{V_R}|<0.33$. 
Nevertheless, it could be relaxed in some LQ models with the mass of the LQ particle to be $M_\text{LQ} \gtrsim 2\,\text{TeV}$ as seen in Table~\ref{tab:LHC_bound_one}.

As for the complex $C_{S_L}$ scenario, it looks that the best fit point in Eq.\,(\ref{eq:fit_CSL_complex}) is disfavored by the LHC and $B_c$ lifetime constraints.
It is noted, however, that the LHC bound is not always proper and depends on the NP model. 
In the case of the charged Higgs model, for instance, the bound on the $s$-channel mediator $H^\pm$ significantly depends on the resonant mass.
Experimentally it is not easy to probe the low mass $\tau\nu$ resonance due to the huge SM $W$ background.
Reference~\cite{Iguro:2022uzz} points out that the range of $m_t\leq m_{H^\pm}\leq 400\,\text{GeV}$ is still viable for the $1\sigma$ explanation, 
although LHC Run\,2 data is already enough to probe this range if the $\tau\nu+b$ signature is searched~\cite{Blanke:2022pjy}. 
Thus, we leave the LHC bound for the complex $C_{S_L}$ scenario below. 
Once the $B_c$ bound of eq.~\eqref{eq:Bc} is imposed, we find
\begin{align}
 & C_{S_L} \simeq -0.58 \pm i\, 0.88 & & \text{Pull} = 4.2 \,,& \label{eq:fit_CSL_complex_withBc}
\end{align}
for the best Pull within the constraint. 

It has been pointed out that $q^2$ distribution in $\Bb \to D^{(\ast)} \tau \overline{\nu}$ is sensitive to the scalar contribution \cite{Sakaki:2014sea,Celis:2016azn}. 
We do not consider the constraint since the detailed correlation among the bins is not available and the constraints depends on $|V_{cb}|$.
Furthermore, experimental analyses for the $q^2$-distribution measurement rely on the theoretical models \cite{Lees:2013uzd,Huschle:2015rga}.
In any case, the Belle II data will be important to resolve these issues~\cite{Kou:2018nap}.

\begin{figure}[t]
\begin{center}
 \includegraphics[width=1 \textwidth]{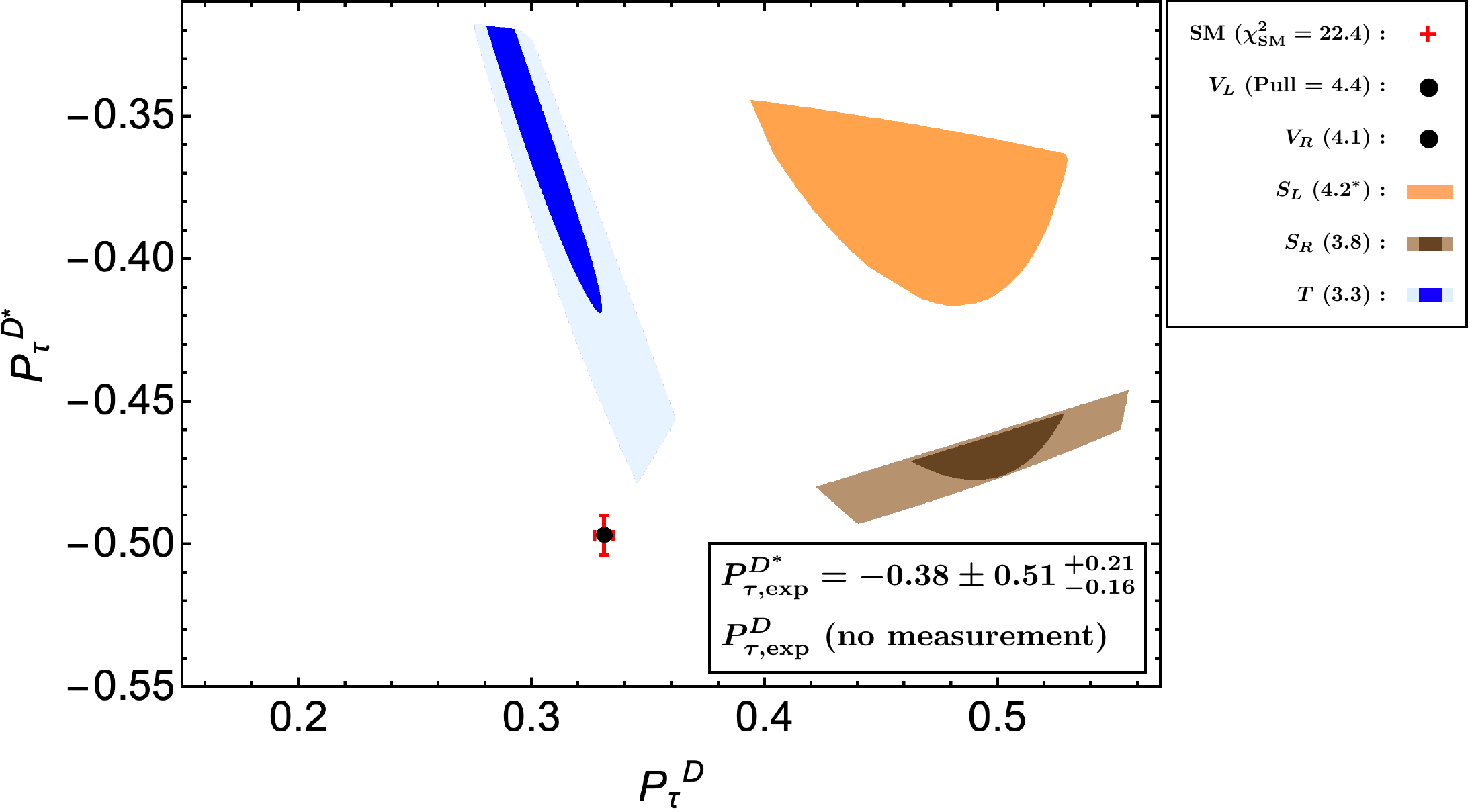}
\end{center}
 \caption{
 \label{fig:EFTprediction}
 Predictions of $P_\tau^D$ and $P_{\tau}^{D^\ast}$ in the single complex NP operator scenarios. 
 The allowed regions satisfying $\Delta\chi^2 \le1$ ($4$)  are shown in dark (light), orange, brown, and blue for the $C_{S_L}$, $C_{S_R}$, and $C_T$ scenarios, respectively, 
 whereas the black dot is the case for the $C_{V_{L,R}}$ scenarios. 
 The $B_c$ lifetime and LHC bounds are also taken into account. 
 As for the $S_L$ scenario, the $B_c$ lifetime rules out the region for $\Delta\chi^2 \le1$, whereas the LHC bound is not taken as discussed in the main text. 
  }
\end{figure}

In Fig.~\ref{fig:EFTprediction}, we show predictions on the plane of $P_\tau^D$--$P_\tau^{D^*}$ from our fit analysis with each complex WC scenario.
The allowed regions satisfying $\Delta\chi^2 \le1$ ($4$) are shown in dark (light) orange, brown, and blue for the complex $C_{S_L}$, $C_{S_R}$, and $C_{T}$ scenarios, respectively, 
where the $B_c$ lifetime and LHC bounds based on the EFT framework are also taken into account.
The $C_{V_{L,\,R}}$ scenarios do not deviate $P_\tau^D$ and $P_\tau^{D^*}$ from the SM predictions as shown with the black dot in the figure. 
Also note that each shaded region is based on different Pull values, implying  different significance, 
in Fig.~\ref{fig:EFTprediction}. 
We can see that the correlation in $\tau$ polarization observables provide the unique predictions that can identify the NP scenarios. 
On the other hand, $R_{\Jpsi}$ is less helpful to distinguish the different operators. 

\subsection{LQ scenarios} 
Finally, we study several LQ scenarios. 
It is well known that three categories of LQs can address the $R_{D^{(*)}}$ anomalies~\cite{Sakaki:2013bfa}, which are referred to as a $SU(2)_L$-singlet vector $\text{U}_1^\mu$, a $SU(2)_L$-singlet scalar $\text{S}_1$, and a $SU(2)_L$-doublet scalar $\text{R}_2$. 
The relevant LQ interactions are given in Appendix~\ref{sec:LQint}.

A key feature with respect to the fit is that these LQ scenarios involve three independent couplings relevant for $b \to c \tau \nu$, 
which are encoded in terms of the two independent (and complex in general) WCs as 
\begin{align}
 \text{U}_1^\mu :~ &  C_{V_L} \,, ~C_{S_R} \,, \\
 \text{S}_1 :~ & C_{V_L} \,, ~C_{S_L} = -4C_T \,, \\
 \text{R}_2 :~ &  C_{V_R} \,, ~C_{S_L} = 4C_T \,,
\end{align}
at the LQ scale $\Lambda_\text{LQ} = M_\text{LQ}$. 
The $SU(2)_L$ doublet vector leptoquark $\text{V}_2$ forms $C_{S_R}$~\cite{Sakaki:2013bfa}, equivalent to the single $C_{S_R}$ scenario, 
and hence this LQ has now the viable solution as seen in Sec.~\ref{sec:singleWCfit}. 
Flavor and collider phenomenologies of $\text{V}_2$ LQ could be interesting, but we leave it for a future work~\cite{Iguro_V2LQ}.

The $C_{V_L}$ phase in $|1+C_{V_L}|^2$ can be absorbed~\cite{Iguro:2018vqb} in the flavor process. 
Thus, the absorption of the $C_{V_L}$ phase is irrelevant for the fit within the flavor observables 
and we take $C_{V_L}$ in $\text{U}_1$ and $\text{S}_1$ LQs as real without loss of generality.\footnote{
Now the real $C_{V_L}$ fit to the $R_{D^{(*)}}$ anomalies gives the minimum $|C_{V_L}|$, and thus is less constrained from the LHC data. 
}
As for $C_{V_R}$ in the $\text{R}_2$ LQ, we assume it as pure imaginary from the fact of Eq.~\eqref{eq:complex_start}.
Therefore, the three LQ scenarios of our interest have three degrees of freedom for the fit and the relevant observables, 
and then it is expected that fit results could be different from the previous studies.

These years, UV completions of the LQ scenarios have been studied in the literature; 
Refs.~\cite{DiLuzio:2017vat,Greljo:2018tuh,Cornella:2019hct,DiLuzio:2018zxy,Bordone:2017bld,Bordone:2018nbg,Blanke:2018sro,Balaji:2018zna,Balaji:2019kwe,Fuentes-Martin:2020bnh,Fuentes-Martin:2020hvc,Guadagnoli:2020tlx,Dolan:2020doe,King:2021jeo,Iguro:2021kdw,Iguro:2022ozl} for $\text{U}_1$, 
Refs.~\cite{Heeck:2018ntp,Marzocca:2018wcf,Marzocca:2021azj} for $\text{S}_1$, 
Refs.~\cite{Becirevic:2018afm,Babu:2020hun} for $\text{R}_2$, 
and see also Refs.~\cite{Faber:2018qon,Faber:2018afz}.
In the next subsection, we consider the case if the $\text{U}_1$ LQ is induced by a UV completed theory that gives a specific relation to the LQ couplings, 
and see how it changes the fit result. 
Recent re-evaluations on mass differences of the neutral $B$ mesons $\Delta M_d,\,\Delta M_s$, 
(improved by HQET sum rule and lattice calculations~\cite{DiLuzio:2019jyq}), 
would constrain a UV-completed TeV-scale LQ model~\cite{Calibbi:2017qbu,Marzocca:2018wcf,DiLuzio:2018zxy,Cornella:2019hct,Crivellin:2019dwb,Iguro:2022ozl}.
In particular, the ratio $\Delta M_d/\Delta M_s$ provides a striking constraint on the coupling texture of the LQ interactions. 
Here, we comment that a typical UV completion requires a vector-like lepton (VLL) and it induces LQ--VLL box diagrams that contribute to $\Delta M_{d,s}$. 
This implies that the constraint of our concern depends on the vector-like fermion mass spectrum, and hence we do not consider $\Delta M_{d,s}$ further in our analysis.

The LQ mass has been directly constrained as $M_\text{LQ} \gtrsim 1.5\,\text{TeV}$ from the LQ pair production searches~\cite{Sirunyan:2018vhk,Aaboud:2019bye,Aad:2021rrh}. 
Hence we take $M_\text{LQ} = 2\,\text{TeV}$ for our benchmark scale. 
We recap that the WCs are bounded from the $\tau + \text{missing}$ search and, 
as shown in Table~\ref{tab:LHC_bound_one}, the LQ scenarios receive milder constraints than the EFT operators as long as $M_\text{LQ} \le 10\,\text{TeV}$.

The WCs will be fitted at the $\mu_b$ scale in our analysis, and then they are related to the WCs defined at the $\Lambda_\text{LQ} = M_\text{LQ}$ scale. 
The renormalization-group equations (RGEs) (the first matrix below)~\cite{Jenkins:2013wua,Alonso:2013hga,Gonzalez-Alonso:2017iyc} and the LQ-charge independent QCD one-loop matching (the second one)~\cite{Aebischer:2018acj} give the following relation
\begin{align}
\begin{pmatrix}
C_{V_L} (\mu_b) \\
C_{V_R} (\mu_b) \\
C_{S_L} (\mu_b) \\
C_{S_R}(\mu_b) \\
C_T (\mu_b)  
\end{pmatrix}
&\simeq
\begin{pmatrix}
1 & 0 &0&0&0\\
0 & 1&0&0&0\\
0 & 0&1.82&0&-0.35\\
0 & 0&0&1.82&0\\
0 & 0&-0.004&0&0.83
\end{pmatrix}
\begin{pmatrix}
1.12 & 0 &0&0&0\\
0 & 1.07&0&0&0\\
0 & 0&1.10&0&0\\
0 & 0&0&1.05&0\\
0 & 0&0&0&1.07
\end{pmatrix}
\begin{pmatrix}
C_{V_L} (\Lambda_\text{LQ} 
) \\
C_{V_R} (\Lambda_\text{LQ} 
)  \\
C_{S_L} (\Lambda_\text{LQ} 
)  \\
C_{S_R} (\Lambda_\text{LQ} 
)  \\
C_T (\Lambda_\text{LQ} 
)   
\end{pmatrix} 
\nonumber
\end{align} 
\begin{align} 
&\simeq 
\begin{pmatrix}
1.12 & 0 &0&0&0\\
0 & 1.07&0&0&0\\
0 & 0&1.91&0&-0.38\\
0 & 0&0&2.00&0\\
0 & 0&0.&0&0.89
\end{pmatrix}
\begin{pmatrix}
C_{V_L} (\Lambda_\text{LQ} 
) \\
C_{V_R} (\Lambda_\text{LQ} 
)  \\
C_{S_L} (\Lambda_\text{LQ} 
)  \\
C_{S_R} (\Lambda_\text{LQ} 
)  \\
C_T (\Lambda_\text{LQ} 
)   
\end{pmatrix} \,, 
\label{eq:RGE2tev}
\end{align} 
with $\Lambda_\text{LQ} = 2\,\text{TeV}$.
Using these numbers, we obtain $C_{S_L}(\mu_b) = -8.9\, C_T(\mu_b)$ and $C_{S_L}(\mu_b) = 8.4\, C_T(\mu_b)$ for $\text{S}_1$ and $\text{R}_2$ LQs, respectively.

With these ingredients, the LQ scenarios in terms of $C_X (\mu_b)$ up to three degrees of freedom are investigated, 
where the full variable case is referred to as the general LQ. 
The results of the best fit points for the general LQ scenarios are then summarized as 
\begin{align}
 \label{eq:LQfit_start}
 &\text{U}_1\text{~LQ}:~& 
 & C_{V_L}=0.07\,,~C_{S_R}= 0.06\,,&
 &\text{Pull} = 3.8\,,&
 \\[0.5em]
 &\text{S}_1\text{~LQ}:~& 
 &C_{V_L}= 0.06\,,~C_{S_L}= -8.9\,C_T = 0.06\,, & 
 &\text{Pull} = 3.8\,,& 
 \\[0.5em]
 \label{eq:R2fit}
 &\text{R}_2\text{~LQ}:~& 
 &C_{V_R}=\pm i 0.68\,,~C_{S_L}= 8.4\,C_T = 0.04 \mp i 0.65\,,&
 &\text{Pull} = 3.8\,.& 
\end{align}
We observe that these three general LQ scenarios have the same Pull which means equivalently favored by the current data.
We also see that $C_{S_{R}}$ ($C_{S_{L}}$) is preferred to be real at the best fit point for the $\text{U}_1$ ($\text{S}_1$) LQ scenario, 
whereas $C_{S_{L}}$ for $\text{R}_2$ LQ is given complex. 
The fit results for $\text{S}_1$ LQ and $\text{R}_2$ LQ with $C_{V_L} =C_{V_R} = 0$ are obtained as 
\begin{align}
 &\text{S}_1\text{~LQ ($C_{V_L} = 0$)}:~& 
 &C_{S_L}= -8.9\,C_T = 0.19\,,&
 &\text{Pull} = 3.9\,,& 
 \\[0.5em]
 &\text{R}_2\text{~LQ ($C_{V_R} = 0$)}:~& 
 &C_{S_L}= 8.4\,C_T = -0.07 \pm i 0.58\,,&
 &\text{Pull} = 4.0\,, 
 \label{eq:LQfit_end}
\end{align}
where the improvements of Pull only come from the benefit of reducing the variables.

In turn, we evaluate the LHC bound on the two independent variables, such as $(C_{V_L},C_{S_R})$, by the following interpretation 
\begin{align}
 &\text{U}_1\text{~LQ}:~~
 \frac{|C_{V_L}(\mu_b)|^2}{(0.42)^2} + \frac{|C_{S_R}(\mu_b)|^2}{(0.77)^2} < 1 \,, & \label{eq:VLSR} \\
 &\text{S}_1\text{~LQ}:~~
 \frac{|C_{V_L}(\mu_b)|^2}{(0.42)^2} + \frac{|C_{S_L}(\mu_b)|^2}{(0.80)^2} < 1 \,, & \label{eq:VLSL} \\
 &\text{R}_2\text{~LQ}:~~
 \frac{|C_{V_R}(\mu_b)|^2}{(0.51)^2} + \frac{|C_{S_L}(\mu_b)|^2}{(0.80)^2} < 1 \,, & \label{eq:VRSL}
\end{align}
where the denominators are the current LHC bounds for the single WC scenarios with $M_\text{LQ} = 2\,\text{TeV}$ from Table~\ref{tab:LHC_bound_one}. 
Indeed this is a good approximation since the bound comes from the high-$p_T$ region that suppresses the interference term between the $V_{L,R}$ and $S_{L,R}$ operators. 
It can be seen that the best fit point of Eq.~\eqref{eq:R2fit} for $\text{R}_2$ LQ is not consistent with the LHC bound of Eq.~\eqref{eq:VRSL}.

\begin{table}[t]
\centering
\newcommand{\bhline}[1]{\noalign{\hrule height #1}}
\renewcommand{\arraystretch}{1.5}
   \begin{adjustbox}{width=\columnwidth,center}
  \begin{tabular}{l c c|c c|c c c} 
  \bhline{1 pt}
 \multirow{2}{*}{ } & 
 \multirow{2}{*}{Pull~[$\chi^2_\text{best}$]} & 
 \multirow{2}{*}{Fitted $C_X$} &
 \multicolumn{2}{c|}{Allowed region of $C_X$}&
 \multicolumn{3}{c}{Predictions $(\Delta\chi^2 \le 1)$} \\
 %
 & & 
 & $B_c \to \tau \overline \nu$ & LHC & $P_\tau^D$ & $-P_\tau^{D^*}$ & $R_{\Jpsi}$  \\
 \hline
 SM &
 --${~[22.4]}$ &
 -- &
 -- &  -- &
 $0.331 \pm 0.004$ &
 $0.497 \pm 0.007$ &
 $0.258 \pm 0.004$ \\
 \hline
 %
 %
 \multirow{3}{*}{$\text{U}_1$ LQ} & 
 \multirow{3}{*}{$3.8~[2.2]$} &
 $~~~\,C_{V_L}:\,[+0.04,+0.10]$ & 
 \multirow{3}{*}{Eq.~\eqref{eq:Bc}} & \multirow{3}{*}{Eq.~\eqref{eq:VLSR}} &
 \multirow{3}{*}{$[0.315, 0.448]$} &
 \multirow{3}{*}{$[0.474, 0.526]$} &
 \multirow{3}{*}{$[0.282, 0.309]$} \\
 & 
 &
 $\text{Re}C_{S_R}:\,[-0.02,+0.14]$ &
 &
 &
 \\
 & 
 &
 $\text{Im}C_{S_R}:\, [-0.46,+0.46]$ &
  &
 &
 \\ \hline
 %
 \multirow{3}{*}{$\text{S}_1$ LQ} & 
 \multirow{3}{*}{$3.8~[2.4]$} &
 $~~~\,C_{V_L}:\,[+0.02,+0.16]$ & 
 \multirow{3}{*}{Eq.~\eqref{eq:Bc}} & \multirow{3}{*}{Eq.~\eqref{eq:VLSL}} &
 \multirow{3}{*}{$[0.104, 0.483]$} &
 \multirow{3}{*}{$[0.373, 0.504]$} &
 \multirow{3}{*}{$[0.269, 0.311]$} \\
 & 
 &
 $\text{Re}C_{S_L}:\,[-0.05,+0.16]$ &
 &
 &
 \\
 & 
 &
 $\text{Im}C_{S_L}:\, [-0.50,+0.50]$ &
  &
 &
 \\ \hline
 %
 \multirow{3}{*}{$\text{R}_2$ LQ} & 
 \multirow{3}{*}{$3.8~[2.0]$} &
 $\text{Im}C_{V_R}:\,[\pm 0.00,\pm 0.50]$ & 
 \multirow{3}{*}{Eq.~\eqref{eq:Bc}} & \multirow{3}{*}{Eq.~\eqref{eq:VRSL}} &
 \multirow{3}{*}{$[0.286, 0.492]$} &
 \multirow{3}{*}{$[0.404, 0.537]$} &
 \multirow{3}{*}{$[0.280, 0.311]$} \\
 & 
 &
 $\text{Re}C_{S_L}:\,[-0.01,+0.08]$ &
 &
 &
 \\
 & 
 &
 $\text{Im}C_{S_L}:\, [\mp 0.00, \mp 0.39]$ &
  &
 &
 \\ \hline
 %
 $\text{R}_2$ LQ &
 \multirow{2}{*}{$4.0~[2.8]$} &
 $\text{Re}C_{S_L}:\,[-0.13,-0.02]$ & 
 \multirow{2}{*}{Eq.~\eqref{eq:Bc}} &  
 \multirow{2}{*}{Eq.~\eqref{eq:VRSL}} &
 \multirow{2}{*}{$[0.426, 0.500]$} &
 \multirow{2}{*}{$[0.407, 0.443]$} &
 \multirow{2}{*}{$[0.276, 0.297]$} \\
 $(C_{V_R}=0)$ & 
 &
 $\text{Im}C_{S_L}:\,[\pm 0.52,\pm 0.65]$ &
 &  &
 &
 \\ \hline
 %
\bhline{1 pt}
 \end{tabular}
 \end{adjustbox}
 \caption{
 The fit results of the $\text{U}_1$, $\text{S}_1$, and $\text{R}_2$ LQ scenarios for $M_\text{LQ} = 2\,\text{TeV}$. 
 The WCs are given at the $\mu_b$ scale. 
 The structure is the same as in Table~\ref{tab:fit:singlereal}. 
 }
 \label{tab:fit:leptoquark}
\end{table}  

In Table~\ref{tab:fit:leptoquark}, we show our fit results and predictions  with respect to the LQ scenarios like we did for the EFT cases. 
It is observed that the general LQ scenarios have less predictive values of the tau polarizations. 
This can be understood from the fact that the complex scalar WCs give large impacts on the interference terms as can be checked from Eqs.~\eqref{eq:PtauD} and \eqref{eq:PtauDs}, 
which result in the wide ranges of the predictions.

\begin{figure}[t]
\begin{center}
 \includegraphics[width=1 \textwidth]{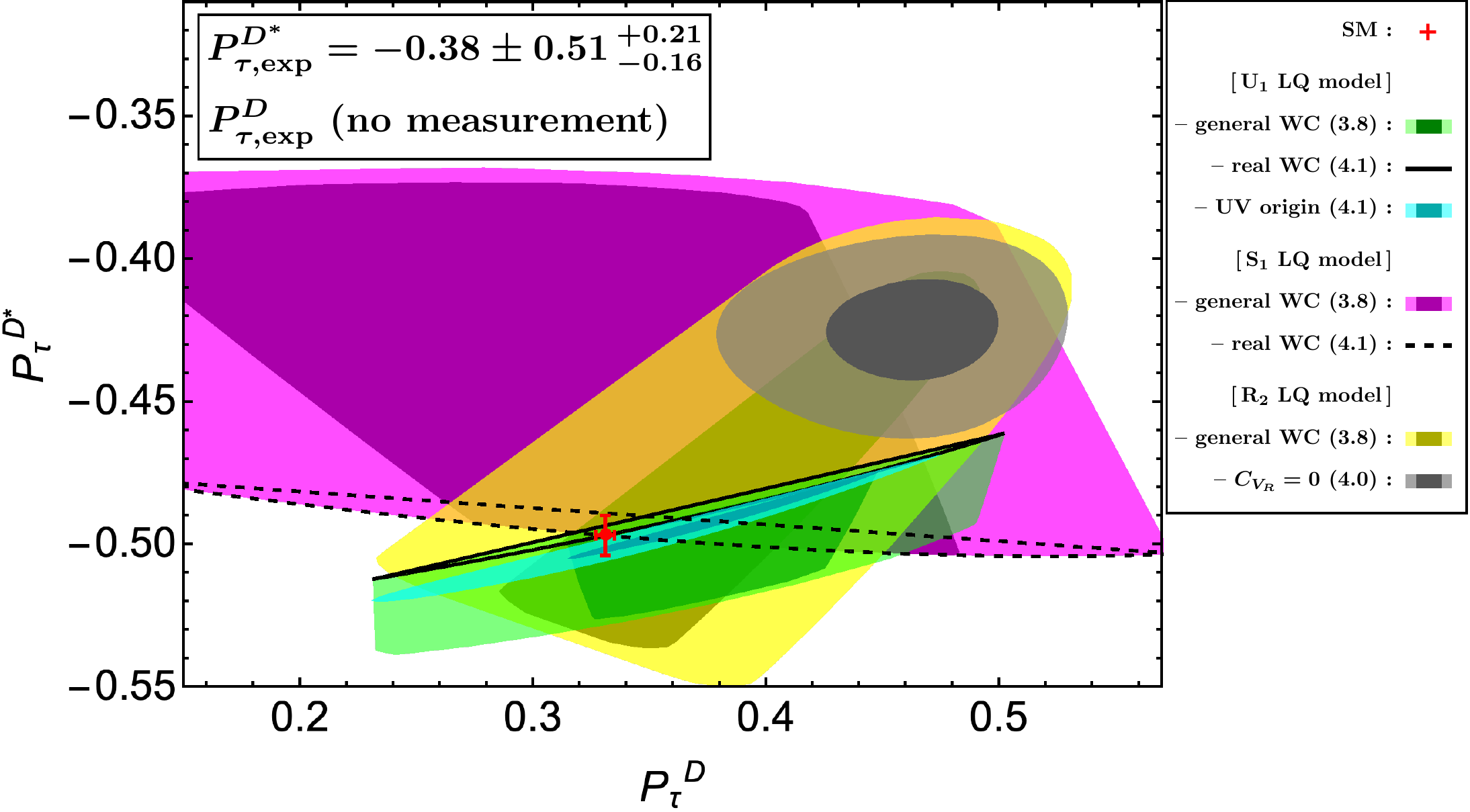}
\end{center}
 \caption{
 Predictions of $P_\tau^D$ and $P_{\tau}^{D^\ast}$ in the LQ scenarios following the same procedure as in Fig.~\ref{fig:EFTprediction}. 
 The allowed regions are shown in dark (light) green, magenta, and yellow for the general $\text{U}_1$, $\text{S}_1$, and $\text{R}_2$ LQ scenarios, respectively. 
 The specific scenarios; 
 $\text{U}_1$ LQ with UV origin (cyan), real WC (solid line);  
 $\text{S}_1$ LQ with real WC (dashed line); and
 $\text{R}_2$ LQ with $C_{V_R}=0$ (gray), are also given.  
  \label{fig:LQprediction}
 }
\end{figure}

Figure~\ref{fig:LQprediction} visualizes the combined $P_\tau^D$--$P_\tau^{D^*}$ predictions satisfying $\Delta\chi^2 \le1\,(4)$ and the aforementioned bounds, 
where the general $\text{U}_1$, $\text{S}_1$, and $\text{R}_2$ LQ scenarios are shown in dark (light) green, magenta, and yellow, respectively. 
The $\text{U}_1$ and $\text{R}_2$ LQ scenarios produce the correlated regions of the $P_\tau^D$--$P_\tau^{D^*}$ predictions and hence could be distinguished. 
On the other hand, the $\text{S}_1$ LQ scenario has the less-predictive wide region, which is hard to be identified.

Figure~\ref{fig:LQprediction} also exhibits the predictions for the several specific scenarios, \ie, 
$\text{U}_1$ LQ with real WC (solid line), $\text{S}_1$ LQ with real WC (dashed line), and $\text{R}_2$ LQ with $C_{V_R}=0$ (gray region).  
It is seen that reducing the variable in the general LQ scenario provides the distinct prediction in particular for $P_\tau^{D^*}$ and 
the correlation for $P_\tau^D$--$P_\tau^{D^*}$ becomes a useful tool to identify the LQ signature.
Therefore, it is significant to restrict the LQ interactions by other processes or by a UV theory that realizes the LQ particle. 
The latter will be discussed in the next subsection for the $\text{U}_1$ LQ (corresponding to the cyan region in the figure). 
Regarding the $\text{S}_1$ LQ scenario,  
we comment that a part of the allowed parameter region is ruled out by the $B \to K^\ast \nu\overline\nu$ measurement and $\Delta M_s$ (via LQ--$\nu_\tau$ box) \cite{Endo:2021lhi}.

\subsection{UV completion of \texorpdfstring{$\text{U}_1$}{U1} LQ}
As the $\text{U}_1$ LQ provides a unique solution, not only to the $b \to c \tau \nu$ anomaly, but also to several flavor issues, UV completions of the $U_1$ LQ have been discussed
~\cite{Barbieri:1995uv,Barbieri:1997tu,Barbieri:2011ci,Barbieri:2011fc,Barbieri:2012uh,Blankenburg:2012nx,Barbieri:2015yvd,Fuentes-Martin:2019mun,FernandezNavarro:2022gst}.  
A typical description is that the $\text{U}_1$ LQ is given as a gauge boson, embedded in a large gauge symmetry, 
such that the third-generation quarks and leptons are coupled to $U_1$ in the interaction basis. 
This means that the two LQ interactions of Eq.~\eqref{eq:U1int} are represented as a universal gauge coupling, $x_L^{33} = x_R^{33} \equiv g_U$ (see Appendix~\ref{sec:LQint}). 
Moving to the mass basis leads to 
\begin{align}
 \label{eq:U2WC_NPscale}
 C_{S_R} (\Lambda_{\text{LQ}})= -2 \beta_R \times C_{V_L} (\Lambda_{\text{LQ}}) \,,     
\end{align}
where $\beta_R = e^{i \phi}$ denotes the relative complex (CP-violating) phase~\cite{Fuentes-Martin:2019mun}, 
which comes from the fact that the phases in the rotation matrices (to the mass basis) for quark and lepton are not necessarily identical. 
The LHC bound for this scenario has been studied and the typical scale of the constraint is obtained as $\Lambda_{\text{LQ}} \gtrsim 3.5\,\text{TeV}$~\cite{Cornella:2019hct}.

The RGE running effect changes the above relation of Eq.~\eqref{eq:U2WC_NPscale} at the $\mu_b$ scale of our interest. 
By taking $\Lambda_{\text{LQ}} = 4\,\text{TeV}$ as a benchmark scale, we obtain 
\begin{align}
\begin{aligned}
C_{V_L} (\mu_b) &= 1\times 1.11 \times C_{V_L} (\Lambda_{\text{LQ}} 
) \,,\\
C_{S_R}(\mu_b) &= 1.90 \times 1.09 \times C_{S_R} (\Lambda_{\text{LQ}}
) \,,
\label{eq:RGE4tev}
\end{aligned}    
\end{align}
where the first coefficient is the QCD two-loop RGE factor~\cite{Gonzalez-Alonso:2017iyc} and the second is the QCD one-loop matching correction~\cite{Aebischer:2018acj} at the NP scale.
Therefore, we have 
\begin{align}
 C_{S_R}(\mu_b) \simeq -3.7\, \beta_R \times C_{V_L}(\mu_b) \,, 
\end{align}
in the case of the UV origin $\text{U}_1$ LQ scenario, applied to our fit analysis.

The result of the best-fit point for the UV origin $\text{U}_1$ LQ scenario, with the definition of $\beta_R = e^{i \phi}$, is shown as 
\begin{align}
\label{eq:U2result}
 & (C_{V_L}, \phi) \,\simeq\, (0.07\,, \pm 0.54\pi) & & \text{Pull} = 4.1\,.&
\end{align}
One can see that this is consistent with the $B_c$ lifetime and LHC bounds. 
Predictions of the observables within $\Delta\chi^2 \le 1,4$ are then given in Fig.~\ref{fig:LQprediction}. 
It is observed that the large complex phase is favored which suppress the interference.
It should be also stressed that the $\tau$ polarizations are so unique that this scenario can be distinguished from the aforementioned LQ scenarios.

\section{The LFU violation in \texorpdfstring{\boldmath{$\Upsilon$}}{Upsilon} decays}
\label{sec:Upsi_decay}

The UV completed NP models 
contributing to  $b \to c \tau \overline\nu$ processes should also bring a related contribution to
$b \overline{b}\to \tau^+ \tau^-$ or $c \overline{c} \to \tau^+ \tau^-$ interactions \cite{Aloni:2017eny, Garcia-Duque:2021qmg,Garcia-Duque:2022tti}.
In this section, we show that
 $\text{U}_1$ and $\text{R}_2$ LQs predict a robust correlation 
between $b \to c \tau \overline\nu$ and $b \overline{b}\to \tau^+ \tau^-$ via the LQ exchange.

A definition of the LFU observable in the $\Upsilon (nS)$ decays is  
\beq
R_{\Upsilon(nS)}\equiv  \frac{\mathcal{B}(\Upsilon(nS) \to \tau^+ \tau^-)}{\mathcal{B}(\Upsilon(nS) \to \ell^+ \ell^-)}\,,
\eeq
with $n=1,2,3$, where $R_{\Upsilon (nS)} \simeq 1$ holds in the SM. 
As for $n\geq 4$, the leptonic branching ratios are significantly suppressed since a $B\Bb$ decay channel is open.%
\footnote{
A novel method for the $n=4$ mode has been proposed in Ref.~\cite{Genon} by using the inclusive di-leptonic channel 
$\Upsilon(4S) \to \ell^\pm \tau^\mp X (\overline\nu \nu)$, 
which could be probed in the Belle II experiment and is directly related to
$\Gamma (b \to X \tau \nu)/\Gamma(b \to X \ell \nu)$.
}
Since the short- and long-distance QCD corrections \cite{Beneke:2014qea} are independent of the lepton mass, they are canceled in this ratio. 
%
One can also discuss the $c \overline{c} \to l^+ l^-$ LFU observable via $\psi(2S)$ decays.
However, we do not consider it because the present experimental error is relatively large.

Recently, the BaBar collaboration has reported a precise result for measurement of  $R_{\Upsilon(3S)}$ \cite{Lees:2020kom}:  
$
R_{\Upsilon(3S)}^{\rm BaBar} = 0.966 \pm 0.008_{\rm stat} \pm 0.014_{\rm syst},
$
where $\ell = \mu$.
Combing a previous measurement by the  CLEO  collaboration  \cite{Besson:2006gj}, 
an average for the $\Upsilon (3S)$ decay is  
\cite{Garcia-Duque:2021qmg}
\beq
R_{\Upsilon(3S)}^{\rm exp} = 0.968 \pm 0.016\,.
\label{eq:Upsilonexp}
\eeq
This value is consistent with the SM prediction \cite{Aloni:2017eny}
\beq
R_{\Upsilon(3S)}^{\rm SM}  = 0.9948 \pm \mathcal{O}(10^{-5})\,,
\eeq 
at the $1.7\sigma$ level.
The SM prediction slightly deviates from $1$ whose leading correction comes from the difference in the phase space factor between the $\tau/\ell$ modes \cite{VanRoyen:1967nq}.
The next-to-leading contribution comes from the QED correction which depends on the lepton mass~\cite{Bardin:1999ak};  
$\delta_{\rm EM} R_{\Upsilon(nS)}=+0.0002$. 
The tree-level $Z$ exchange also contributes, but its effect is $\mathcal{O}(10^{-5})$ \cite{Aloni:2017eny}.
There is no Higgs boson contribution as one can see below.
The other channels ($n=1,2$) still suffer from
the current experimental uncertainty,
and we do not utilize them in our presentation.

The effective Hamiltonian which is relevant to the bottomonium decay into $\tau^+ \tau^-$ is described as
\beq
- \mathcal{H}^{\rm NP}_{\rm eff} = & ~ C_{VLL}^{b \tau} (\overline{b}\gamma^{\mu}P_L b) (\overline{\tau}\gamma_{\mu} P_L \tau)
+ C_{VRR}^{b \tau} (\overline{b}\gamma^{\mu}P_R b) 
(\overline{\tau}\gamma_{\mu} P_R \tau) \nonumber \\
 & + C_{VLR}^{b \tau} (\overline{b}\gamma^{\mu}P_L b) (\overline{\tau}\gamma_{\mu} P_R \tau)
+C_{VRL}^{b \tau} (\overline{b}\gamma^{\mu}P_R b) (\overline{\tau}\gamma_{\mu} P_L \tau) \\
& + \left[ C_T^{b \tau} (\overline{b} \sigma^{\mu \nu} P_R b )(\overline{\tau} \sigma_{\mu \nu} P_R \tau ) 
+ C_{SL}^{b \tau}  (\overline{b}P_L b)(\overline{\tau}P_L \tau)
+ C_{SR}^{b \tau}  (\overline{b}P_R b)(\overline{\tau}P_L \tau)
+ \textrm{h.c.}\right]\,,
\nonumber 
\eeq
at the scale $\mu = m_{\Upsilon}$. Note that $C_{VLL}^{b \tau}, C_{VRR}^{b \tau}, C_{VLR}^{b \tau}$ and $C_{VRL}^{b \tau}$ are real coefficients, and  
$C_{SL}^{b \tau}$ and $C_{SR}^{b \tau} $ never contribute to the $\Upsilon(nS) \to \tau^+\tau^-$ due to $\langle 0 | \overline{b} b | \Upsilon \rangle =\langle 0 | \overline{b}\gamma_5 b | \Upsilon \rangle=0$.
In this convention, the partial decay width is given by \cite{Aloni:2017eny}
\beq
\begin{aligned}
\Gamma(\Upsilon(nS) \to \tau^+ \tau^-)= &~
\frac{f_\Upsilon^2}{4 \pi m_\Upsilon}
\sqrt{1 - 4 x_{\tau}^2} \Bigl[ 
 A_\Upsilon^2 (1 + 2 x_{\tau}^2) 
+ B_\Upsilon^2 (1 - 4 x_{\tau}^2) 
  \\
& \quad + \frac{1}{2}C_\Upsilon^2  
(1 - 4 x_{\tau}^2)^2 + \frac{1}{2} D_\Upsilon ^2  
(1 - 4 x_{\tau}^2)  + 2   A_\Upsilon 
C_\Upsilon x_{\tau} (1 - 4 x_{\tau}^2)\Bigr]\,,
\end{aligned}
\eeq
with
\beq
A_\Upsilon =& \frac{4 \pi \alpha}{3} + 
\frac{m_{\Upsilon}^2}{4} \left[  C_{VLL}^{b \tau}+  C_{VRR}^{b \tau}+ C_{VLR}^{b \tau}+C_{VRL}^{b \tau}+ 16  x_{\tau} \frac{f_\Upsilon^T}{f_\Upsilon} \textrm{Re}\left(C_T^{b\tau}\right)\right]\,,\\
B_\Upsilon =& \frac{m_{\Upsilon}^2}{4} 
 \left(  C_{VRR}^{b \tau}+ C_{VLR}^{b \tau} -C_{VLL}^{b \tau} -C_{VRL}^{b \tau}\right)\,,\\
C_\Upsilon = & 2 m_\Upsilon^2 \frac{f_\Upsilon^T}{f_\Upsilon}
\textrm{Re} \left( C_T^{b \tau} \right)\,,\\
D_\Upsilon =& 2 m_\Upsilon^2 \frac{f_\Upsilon^T}{f_\Upsilon} 
\textrm{Im} \left( C_T^{b\tau} \right)\,,
\eeq
and 
\beq
x_\tau =\frac{m_{\tau}}{m_{\Upsilon}}\,.
\eeq
The $f_{\Upsilon}$ and $f_{\Upsilon}^T$ are form factors for vector and tensor currents in $\Upsilon$ hadronic-matrix elements, 
and $f_{\Upsilon} = f_{\Upsilon}^T$ holds in the heavy quark limit, which is realized for the $\Upsilon$ decays \cite{Aloni:2017eny}.

Within the SM, this process is predominantly caused by the QED.
Nevertheless, the photon-exchange QED contribution is suppressed by $1/m_{\Upsilon}^2 $, 
and hence the NP contribution could be non-negligible~\cite{Aloni:2017eny,Matsuzaki:2018jui,Garcia-Duque:2021qmg}.
In the SM, 
$A_\Upsilon \simeq 4 \pi \alpha/3$ and $B_\Upsilon, C_\Upsilon, D_\Upsilon \simeq 0$. 
Setting the light lepton mass to zero and $m_\Upsilon = m_{\Upsilon(3S)} = 10.355\,\text{GeV}$, 
we obtain the following numerical formula 
\beq
\begin{aligned}
\frac{R_{\Upsilon(3S)}}{R_{\Upsilon(3S)}^{\rm SM}} =&~  
1 +1.64\times 10^{-3}\ {\rm TeV}^2 \left( C_{VLL}^{b \tau}+  C_{VRR}^{b \tau}+ C_{VLR}^{b \tau}+C_{VRL}^{b \tau}\right) \\
& + 6.37\times 10^{-3}\ {\rm TeV}^2\  \textrm{Re}\left(C_T^{b \tau} \right)+ \delta_\Upsilon\,, 
\end{aligned}
\eeq
with
\beq
\begin{aligned}
\delta_\Upsilon = &~ 5.22 \times 10^{-6}\ {\rm TeV}^4 \left( C_{VLL}^{b \tau}+  C_{VRR}^{b \tau}+ C_{VLR}^{b \tau}+C_{VRL}^{b \tau}\right)\textrm{Re}\left(C_T^{b \tau} \right)
 \\
&+ 6.71 \times 10^{-7}\ {\rm TeV}^4 \left( C_{VLL}^{b \tau}+  C_{VRR}^{b \tau}+ C_{VLR}^{b \tau}+C_{VRL}^{b \tau}\right)^2  
\\
&+5.59 \times 10^{-7}\ {\rm TeV}^4 \left(  C_{VRR}^{b \tau}+ C_{VLR}^{b \tau} -C_{VLL}^{b \tau} -C_{VRL}^{b \tau}\right)^2 
 \\
& + 2.51 \times 10^{-5}\ {\rm TeV}^4 \left[\textrm{Re}\left(C_T^{b \tau} \right)\right]^2  
+  1.79 \times 10^{-5}\ {\rm TeV}^4 \left[\textrm{Im}\left(C_T^{b \tau} \right)\right]^2\,,
\end{aligned}
\eeq
where the $\delta_\Upsilon$ term gives negligible contributions.

Let us now look into a correlation between 
$R_{\Upsilon(3S)}$ and $R_{D^{(\ast)}}$ by using the specific examples of the $\text{U}_1$ and $\text{R}_2$ LQs.
First, we exhibit the $\text{U}_1$ LQ case.
The $\text{U}_1$ LQ interaction with the SM fermions is given in Eq.~\eqref{eq:U1int}.
Integrating the $\text{U}_1$ LQ out, 
as well as the charged current contributions ($b \to c \tau \overline\nu$) in Eq.~\eqref{eq:U1charged}, 
the neutral current ones ($b \overline{b} \to \tau^+ \tau^-$) are  obtained as 
\beq
C_{VLL}^{b\tau}(\mu_{\rm LQ}) 
= -\frac{|x_L^{b\tau}|^2}{m_{U_1}^2}\,,
\quad
C_{VRR}^{b\tau}(\mu_{\rm LQ}) 
= -\frac{|x_R^{b\tau}|^2}{m_{U_1}^2}\,,
\quad
C_{SR}^{b \tau} (\mu_{\rm LQ}) 
= \frac{2 x_L^{b \tau} (x_R^{b \tau})^\ast}{m_{U_1}^2}\,.
\eeq
The vector contributions do not change under the RGEs, while the scalar contribution does not affect the $\Upsilon$ decay.
Here,
an important point is that $R_{\Upsilon(3S)} / R_{\Upsilon(3S)}^{\rm SM}$ 
is predicted to be less than $1$ when NP contributions are dominated by vector interactions.
It would lead to a coherent deviation with $R_{D^{(\ast)}}$.
\begin{figure}[t]
    \centering
    \includegraphics[width=0.7\textwidth]{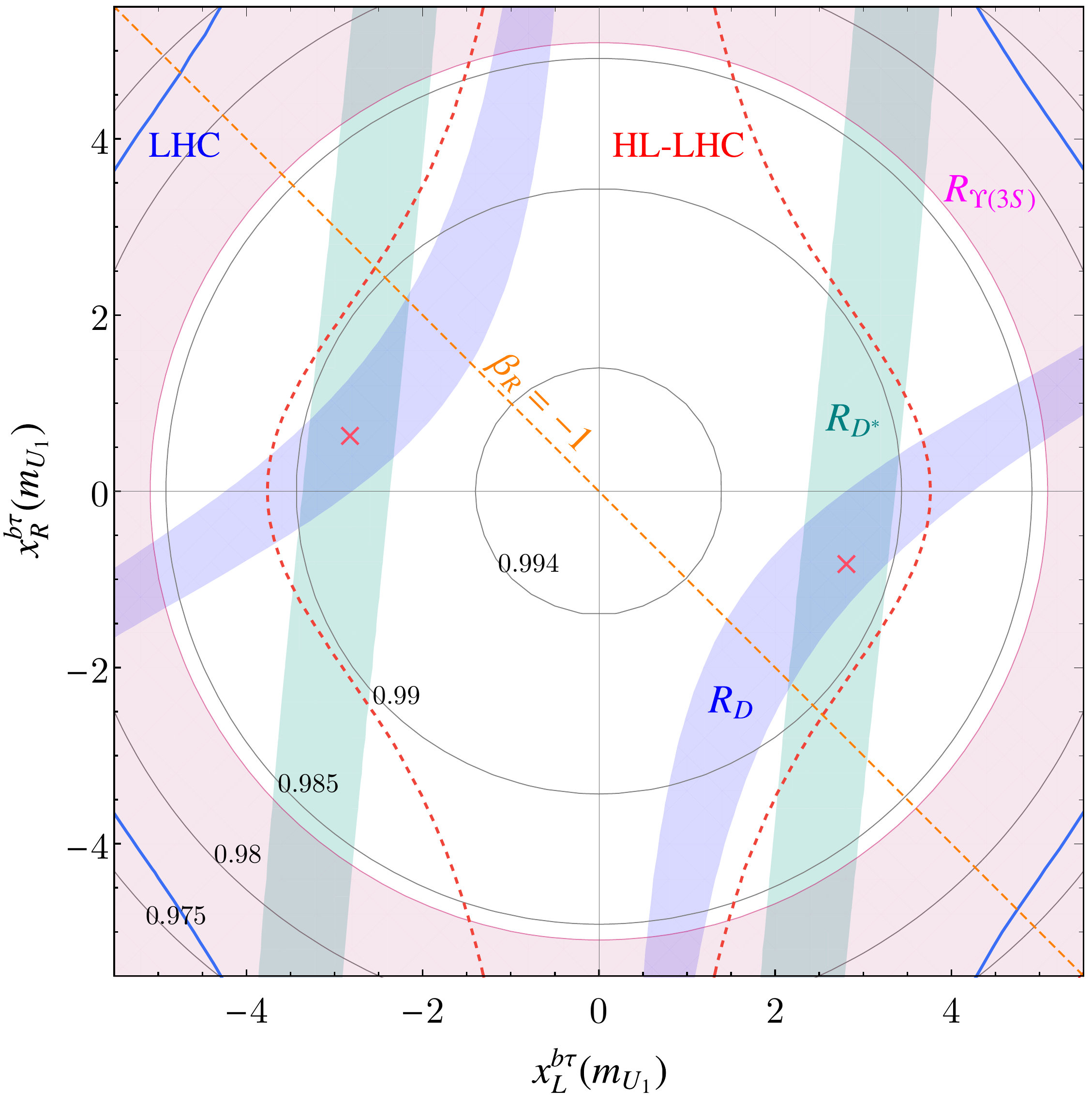}
    \caption{The correlation between 
$R_{\Upsilon(3S)}$ and $R_{D^{(\ast)}}$ is exhibited in the 
$\text{U}_1$ LQ scenario with $m_{\text{U}_1}=2\,$TeV by setting $ (V x_L)^{c \tau}  = V_{cb} x_L^{b \tau}$.
    The predicted values of $R_{\Upsilon(3S)}$ are shown by black contours, and 
    $R_{\Upsilon(3S)}^{\rm exp}$ in Eq.~\eqref{eq:Upsilonexp} is shown in the red shaded area.
    $R_D$ and $R_{D^\ast}$ anomalies can be explained in the blue and green regions, respectively.
    The LHC exclusion region (outside the blue line) and the HL-LHC sensitivity (red dashed line) are based on the result of Table~\ref{tab:LHC_bound_one}.
    The best fit points of Eq.~\eqref{eq:LQfit_start} are shown by red crosses. The orange dashed line represents the $U(2)$ flavor symmetry prediction with $\beta_R = -1$.}
    \label{fig:U1Upsilon}
\end{figure}

Setting $m_{U_1}=2.0\,\text{TeV}$ and $ (V x_L (\mu_{\text{LQ}}))^{c \tau}  = V_{cb} x_L^{b \tau}(\mu_{\text{LQ}})$, namely $x_L^{s \tau}(\mu_{\text{LQ}}) =0$, 
we show a correlation between $R_{\Upsilon(3S)}$ and $R_{D^{(\ast)}}$ in Fig.~\ref{fig:U1Upsilon}. 
Here, favored parameter regions in the $\text{U}_1$ LQ model are exhibited on $x_L^{b\tau}$--$x_R^{b\tau}$ plane at the renormalization scale $\mu_{\text{LQ}} = m_{U_1}$.
The black contour represents the expected values of 
$R_{\Upsilon(3S)}$.
The red shaded region 
is favored by $R_{\Upsilon(3S)}^{\rm exp}$ in Eq.~\eqref{eq:Upsilonexp}.
It is noted that if we adopt the $2\sigma$ constraint of $R_{\Upsilon(3S)}^{\rm exp}$,
the entire parameter region is allowed.
The blue and green regions can explain the $R_D$ and $R_{D^\ast}$ discrepancies 
within $1\sigma$, respectively.
The exclusion region by the LHC analysis ($\tau +\,$missing search) is outside the blue line, 
while the future prospect of the High Luminosity LHC (HL-LHC) is shown by the red dashed line,
see Table~\ref{tab:LHC_bound_one}.\footnote{%
Note that a stronger collider bound would come from a non-resonant $\tau\tau$ search~\cite{ATLAS:2020zms,CMS:2022goy}, 
although it is model-parameter dependent. 
}
Furthermore, the orange dashed line stands for a prediction in the case of the UV origin $\text{U}_1$ LQ with $\beta_R = -1 ~(\phi=\pi)$. 
From the figure, it is found that the current $R_{\Upsilon(3S)}^{\rm exp}$ overshoots favored parameter region from the $R_{D^{(*)}}$ anomalies.
The best fit points of Eq.~\eqref{eq:LQfit_start} are shown by red crosses and predict $R_{\Upsilon(3S)}=0.991$, 
distinct from the 0.99 contour line in the figure. 
Thus, it seems crucial to measure $R_{\Upsilon(nS)}$ with less than $1\%$ accuracy in order to distinguish the $\text{U}_1$ LQ signal.

\begin{figure}[t]
    \centering
    \includegraphics[width=0.7\textwidth]{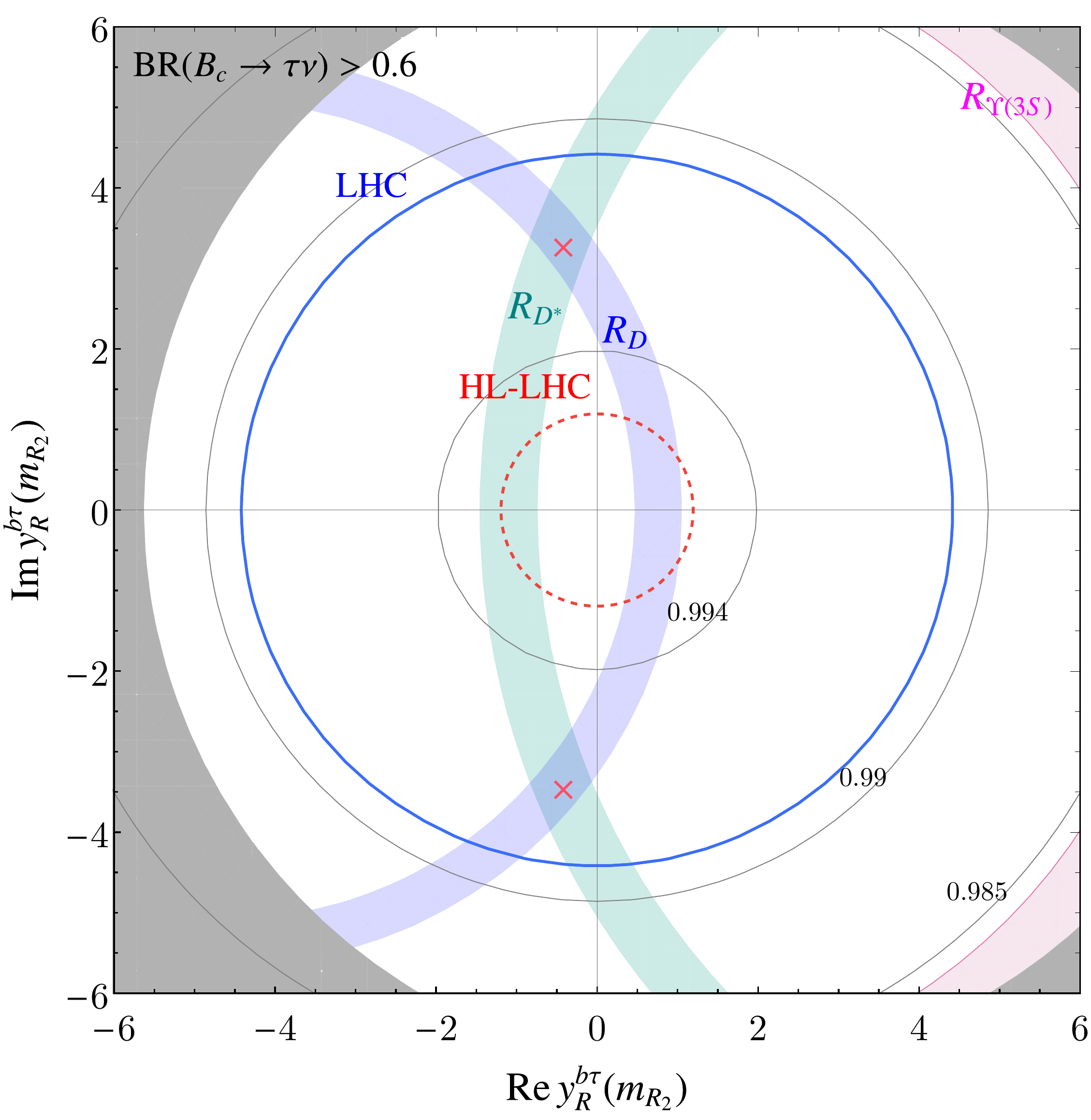}
    \caption{
    Correlation between $R_{\Upsilon(3S)}$ and $R_{D^{(\ast)}}$ in the $\text{R}_2$ LQ scenario with $m_{\text{R}_2}=2\,$TeV and $y_L^{c \tau}=1$.
    The color convention is the same as in Fig.~\ref{fig:U1Upsilon}. 
    The gray shaded region is excluded by the $B_c$ lifetime. 
    The best fit points of Eq.~\eqref{eq:LQfit_end} are shown by red crosses.
\label{fig:R2Upsilon}}
\end{figure}

Next, we investigate the  $\text{R}_2$ LQ scenario.
The $\text{R}_2$ LQ interaction with the SM fermions is given in Eq.~\eqref{eq:R2int}.
The generated charged current contributions are given in Eq.~\eqref{eq:R2charged}, while the neutral current one is
\beq
C_{VLR}^{b \tau} (\mu_{\rm LQ})= - \frac{|y_R^{b \tau}|^2}{2 m_{R_2}^2} \,.
\eeq
Since $C_{VLR}^{b \tau} < 0$, 
$R_{\Upsilon(3S)} / R_{\Upsilon(3S)}^{\rm SM}$ 
has to be less than $1$ again.

The result is shown in Fig.~\ref{fig:R2Upsilon}.
Here, we set $y_L^{c \tau} (\mu_{\rm LQ})=1$, $|V_{cb}|=0.04$, and $m_{\text{R}_2}=2.0\,$TeV, and take $y_R^{b \tau}(\mu_{\rm LQ})$ as complex value.
The color convention is the same as the $\text{U}_1$ LQ case.
Furthermore, the  gray shaded region is  excluded by the $B_c$ lifetime, \ie, $\mathcal{B}(B_c \to \tau\overline\nu) > 0.6$.
The best fit points in Eq.~\eqref{eq:LQfit_end} are shown by red crosses,  predicting $R_{\Upsilon(3S)}=0.992$.
Similar to $\text{U}_1$ LQ interpretation, $1\%$ accuracy of the $R_{\Upsilon(nS)}$ measurement is required as the $\text{R}_2$ LQ signature.

At the current stage, 
the large experimental uncertainty in $R_{\Upsilon(3S)}$ cannot allow a clear-cut conclusion.
One should note that the Belle and Belle II experiments have enough sensitivities to
the $R_{\Upsilon(nS)}$ measurements which would be more accurate than the existing BaBar measurement~\cite{ishikawa}.

\section{Conclusions and discussion}
\label{sec:conclusion}

\begin{table}[t]
\centering
\newcommand{\bhline}[1]{\noalign{\hrule height #1}}
\renewcommand{\arraystretch}{1.5}
\rowcolors{2}{gray!15}{white}
   \scalebox{0.95}{
  \begin{tabular}{lccc ccc cc} 
  \bhline{1 pt}
  \rowcolor{white}
 &Spin & Charge & Operators&$R_D$ &$R_{D^*}$ 
 & LHC & Flavor 
 \\   
  \hline
 $H^\pm$ 
 &0&($\bf{1}$,\,$\bf{2}$,\,$\sfrac{1}{2}$)& $O_{S_L}$& ${\Large{\color{teal}{\bf{\checkmark}}}}$ & ${\Large{\color{teal}{\bf{\checkmark}}}}$ & $b\tau\nu$ &$B_c\to\tau\nu$, $F_L^{D^\ast}$, $P^D_\tau$, $M_W$\\
 $\text{S}_1$
 &0&($\bf{\bar{3}}$,\,$\bf{1}$,\,$\sfrac{1}{3}$)& $O_{V_L}$, $O_{S_L}$, $O_{T}$& ${\Large{\color{teal}{\bf{\checkmark}}}}$ & ${\Large{\color{teal}{\bf{\checkmark}}}}$& $\tau\tau$ & $\Delta M_s$, $P_{\tau}^D$, $B \to K^{(\ast)} \nu \nu $\\
 $\text{R}_2^{(\sfrac{2}{3})}$ &0&($\bf{3}$,\,$\bf{2}$,\,$\sfrac{7}{6}$)&$O_{S_L}$, $O_{T}$, ($O_{V_R}$)& ${\Large{\color{teal}{\bf{\checkmark}}}}$ & ${\Large{\color{teal}{\bf{\checkmark}}}}$ & $b\tau\nu$, $\tau\tau$ &$R_{\Upsilon(nS)}$, $P_{\tau}^{D^\ast}$, $M_W$\\
 $\text{U}_1$ &1&($\bf{3}$,\,$\bf{1}$,\,$\sfrac{2}{3}$)&$O_{V_L}$, $O_{S_R}$& ${\Large{\color{teal}{\bf{\checkmark}}}}$ & ${\Large{\color{teal}{\bf{\checkmark}}}}$& $b\tau\nu$, $\tau \tau$ & $R_{K^{(\ast)}}$, $R_{\Upsilon(nS)}$, $B_s \to \tau\tau$\\
 $\text{V}_2^{(\sfrac{1}{3})}$ &1&($\bf{\bar{3}}$,\,$\bf{2}$,\,$\sfrac{5}{6}$)&$O_{S_R}$& ${\Large{\color{teal}{\bf{\checkmark}}}}$ & ${2\sigma}$& $\tau\tau$ & $B_s \to \tau\tau$, $M_W$\\
\bhline{1 pt}
   \end{tabular}
   }
    \caption{\label{tab:summary_table} 
Summary table for the single-mediator NP scenarios in light of the $b \to c \tau \nu$ anomaly.
We add implications for the LHC searches and flavor observables in the last two columns, which is useful to identify the NP scenario. 
In the $\text{V}_2^{(\sfrac{1}{3})}$ LQ scenario, $2\sigma$ for $R_{D^\ast}$ implies that it can explain the $R_{D^\ast}$ anomaly within the $2\sigma$ range 
(but not within $1\sigma$).
}
\end{table}

In this work we revisited our previous phenomenological investigation and presented a statistical analysis of the LFU violation in $R_{D^{(*)}}$, 
including the new experimental data from the LHCb experiment. 
Starting with the re-evaluation of the generic formulae for $R_{D^{(*)}}$ by employing the recent development of the $B\to D^{(\ast)}$ transition form factors, 
we examined the new physics possibility with the low-energy effective Lagrangian as well as the leptoquark models.
In addition to the constraints from the low-energy observables and the high-$p_T$ mono-$\tau$ search at LHC, 
the predictions on the relevant observables of $R_\Upsilon$, $R_{\Jpsi}$, and the tau polarizations $P_{\tau}^{D^{(*)}}$ are evaluated.

To be precise, we performed the $\chi^2$ fit to the experimental measurements of $R_{D^{(*)}}$ and the $D^*$ polarization $F_L^{D^*}$.
This updated analysis shows that the present data deviates from the SM predictions at $\sim 4\sigma$ level. 
Our fit result is summarized in Table~\ref{tab:fit:singlereal} with Eqs.~\eqref{eq:complex_start}--\eqref{eq:complex_end} for the single-operator scenarios, 
and Table~\ref{tab:fit:leptoquark} with Eqs.~\eqref{eq:LQfit_start}--\eqref{eq:LQfit_end}, \eqref{eq:U2result} for the single-mediator leptoquark scenarios. 
The NP fit improvements compared with the SM one are visualized by Pull as usual, and it was found that the SM-like vector operator still gives the best Pull.

Due to the new LHCb result, the experimental world average has slightly come close to the SM predictions of $R_D$ and $R_{D^*}$. 
This change has affected the previous conclusions such that the scalar NP solutions to the $b \to c \tau \nu$ anomaly had been disfavored. 
Namely, the scalar NP interpretations have been revived now. 
On the other hand, it is found that the results of the LQ scenarios do not drastically change, compared with the previous fit.

As it was pointed out in the literature, 
the precise measurements of the polarization observables $P_{\tau}^{D^{(*)}}$ and $F_L^{D^*}$ have the potential to distinguish the NP scenarios. 
In Figs.~\ref{fig:EFTprediction} and \ref{fig:LQprediction}, we show our predictions of $P_{\tau}^{D}$ and $P_{\tau}^{D^{*}}$ for the possible NP scenarios.
One can make sure that the single-operator NP scenario explaining the $b \to c \tau \nu$ anomaly can be identified by the $P_{\tau}^{D^{(*)}}$ measurements, 
which may be available at the Belle~II experiment.
On the other hand, the general LQ scenarios are hard to be distinguished due to predicting wide ranges of $P_{\tau}^{D^{(*)}}$. 
Once the LQ model with restricted interactions is constructed, however, 
we see that the $P_{\tau}^{D^{(*)}}$ measurement has significant potential to probe the LQ signature.
The high energy collider search is also important since the high-$p_T$ lepton search at the LHC can directly probe the NP interactions affecting the LFU ratios.

We also investigated the NP impacts on the LFU violation in the $\Upsilon(nS)$ decays. 
We found that the LFU ratio $R_\Upsilon$ has to be correlated to $R_{D^{(\ast)}}$ in the $\text{U}_1$ and $\text{R}_2$ LQ scenarios, 
while no correlation is expected in the $\text{S}_1$ LQ scenario. 
It is shown that an experimental accuracy of less than $1\%$ for the $R_{\Upsilon(nS)}$ measurement is necessary in order to identify the LQ scenario. 
We expect that this is possible in the Belle II experiment.

In Table~\ref{tab:summary_table}, we put a summary check sheet to find which single-mediator NP scenarios are viable 
and to see important observables in order to identify the NP scenario responsible for the $b \to c \tau \nu$ anomaly. 

Furthermore, it is known that the baryonic counterpart of $R_{D^{(*)}}$, 
namely $R_{\Lambda_c} \equiv  \mathcal{B}(\Lambda_b \to \Lambda_c \, \tau \overline \nu) / \mathcal{B}(\Lambda_b \to \Lambda_c \,\mu \overline \nu)$, 
provides the independent cross check of the $b \to c \tau \nu$ anomaly~\cite{Blanke:2018yud,Blanke:2019qrx}.
Recently, $R_{\Lambda_c}$ has been measured for the first time as $R_{\Lambda_c} = 0.242 \pm 0.026 \pm 0.040 \pm 0.059$ by the LHCb experiment~\cite{LHCb:2022piu}, 
where the last dominant uncertainty comes from an external branching fraction from the LEP measurement~\cite{DELPHI:2003qft}. 
This result implies consistency with the SM prediction at $1.1\sigma$ level~\cite{Bernlochner:2018kxh,Bernlochner:2018bfn}, 
while, instead, normalizing with the SM prediction of $\Gamma(\Lambda_b\to\Lambda_c\mu\overline{\nu})$ improves the accuracy and slightly up-lifts the central value, 
\eg, $R_{\Lambda_c} = |0.04/V_{cb}|^2 (0.285 \pm 0.073)$~\cite{Bernlochner:2022hyz}.
Even though the current experimental uncertainty is not enough precise, 
it could already provide a nontrivial constraint on the NP parameter space which can explain the $b \to c \tau \nu$ anomaly. 
An implication of the measured $R_{\Lambda_c}$ for NP models is given in Ref.~\cite{Fedele:2022iib}.

\section*{Acknowledgements}
The authors would like to thank 
 Motoi Endo, 
 Akimasa Ishikawa,
 Satoshi Mishima,
 Yuta Takahashi,
 and
 Kei Yamamoto
 for fruitful comments and valuable discussion at different stage of the work.
We also appreciate 
Monika Blanke,
Andreas Crivellin, 
Marco Fedele, 
Ulrich Nierste 
and 
Felix Wilsch for useful discussion.
S.\,I. enjoys the support from the Deutsche Forschungsgemeinschaft (DFG, German Research Foundation) under grant 396021762-TRR\,257.
S.\,I. thanks Karlsruhe House of Young Scientists (KHYS) for the financial support which enabled him to invite R.W. for the discussion.
The work of T.K.\ is supported by the Japan Society for the Promotion of Science (JSPS) Grant-in-Aid for Early-Career Scientists (Grant No.\,19K14706) and the JSPS Core-to-Core Program (Grant No.\,JPJSCCA20200002). 
R.W.\ is partially supported by the INFN grant `FLAVOR' and the PRIN 2017L5W2PT.


\appendix
\section{Leptoquark interactions}
\label{sec:LQint}
The LQ interactions are classified with the generic $SU(3)_c \times SU(2)_L \times U(1)_Y$ invariant form \cite{Buchmuller:1986zs}. 
We leave details of the model constructions, and then just introduce the interactions relevant for $b\to c\tau\overline\nu$.
As mentioned above, there are three viable candidates of leptoquark 
$\text{U}_1,~\text{S}_1,~ \text{R}_2 $\cite{Angelescu:2018tyl}.
Their quantum numbers under $SU(3)_{\rm{C}}$, $SU(2)_{\rm{L}}$, $U(1)_{\rm{Y}}$ are summarized in Table \ref{tab:summary_table}.

First, the $\text{U}_1$ vector LQ interaction with the SM fermions, defined in the interaction basis, is given by 
\beq
\mathcal{L}_{U_1} = \hat x^{ij}_L \overline{Q}_i \gamma_{\mu} U^{\mu}_1 L_j 
+ \hat x^{ij}_R \overline{d}_{Ri}\gamma_{\mu} U^{\mu}_1 \ell_{R j} 
+\textrm{h.c.}\,.
\label{eq:U1int}
\eeq
Integrating out the $\text{U}_1$ LQ mediator particle, then, the Wilson coefficients (WCs) for the charged current of our interest ($b \to c \tau \overline\nu$) is obtained as 
\beq
\begin{aligned}
C_{V_L} (\mu_{\rm LQ}) 
&=
\frac{1}{2 \sqrt{2} G_F V_{cb} } \frac{ (V x_L)^{c \tau} (x_L^{b \tau})^\ast}{m_{U_1}^2}\,,
\\
C_{S_R} (\mu_{\rm LQ}) &= 
-
\frac{1}{\sqrt{2} G_F V_{cb}}\frac{(V x_L)^{c \tau} (x_R^{b \tau})^\ast}{m_{U_1}^2}\,,
\label{eq:U1charged}
\end{aligned}
\eeq
where $V$ is the CKM matrix and the couplings $x_{L,R}$ are in the mass basis. 
The relative sign and factor two in Eq.~\eqref{eq:U1charged} come from the property of Fierz identity.

In a typical UV completed theory~\cite{Fuentes-Martin:2019mun}, 
the $\text{U}_1$ LQ is realized as a gauge boson generated from a large gauge symmetry and only couples to the third-generation SM fermions. 
Namely, $\hat x_R^{b \tau} = \hat x_L^{b \tau} \equiv g_U$, with the others to be zero, is indicated in the gauge interaction basis. 
Moving to the mass basis, then, generates a non-zero off-diagonal part such as $x_L^{c \tau}$ and also $x_R^{b \tau} = e^{i \phi} x_L^{b \tau}$, 
where the phase comes from those in the rotation matrices to the mass bases of the left- and right-handed quark and lepton fields that are not canceled in general. 
Therefore, the UV completion of $\text{U}_1$ LQ suggests 
\begin{align}
    C_{S_R}  (\mu_{\rm LQ}) =  - 2 e^{i \phi} C_{V_L}  (\mu_{\rm LQ})\,, 
\end{align} 
as introduced in the main text. 
We also comment that an extension of the fermion families with a nontrivial texture of the fermion mass matrices is necessary to construct a practical UV model~\cite{Iguro:2021kdw}.


The $\text{S}_1$ scalar LQ interaction in the mass basis is given by  
\beq
\mathcal{L}_{{\rm S}_1} 
&=  \big{(}V^\ast y_L \big{)}^{ij}\, \overline{u^C_{L\,i}}\ell_{L\,j}{\rm S}_1-y_L^{ij}\,\overline{d^C_{L\,i}}\nu_{L\,j}{\rm S}_1+y_R^{ij}\, \overline{u^C_{R\,i}}\ell_{R\,j}  {\rm S}_1+ \mathrm{h.c.}\,.
\eeq
In the scalar LQ scenario, the source of the generation violating couplings is off-diagonal element of  Yukawa matrices.  
Then the four-fermion interactions of $b \to c \tau {\overline \nu}$ are given by 
\beq
\begin{aligned}
C_{S_L} (\mu_{\rm LQ})
&= - 4 \, C_T (\mu_{\rm LQ})
= -\dfrac{1}{4 \sqrt{2} G_F V_{cb}}\dfrac{y_L^{b{\tau}} \big{(}y_R^{c\tau}\big{)}^\ast}{ m_{S_1}^2}\,, \\
C_{V_L} (\mu_{\rm LQ})
&= \dfrac{1}{4 \sqrt{2} G_F V_{cb}}\dfrac{y_L^{b \tau}\big{(}V y_L^\ast\big{)}^{c\tau}}{ m_{S_1}^2} \,.
\end{aligned}
\eeq

Finally, we introduce the $\text{R}_2$ scalar LQ interaction.
$\text{R}_2$ is a SU(2) doublet and a component with $2/3$ of the electromagnetic charge ${\rm R}_2^{(\sfrac{2}{3})}$ can contribute to $b \to c \tau {\overline \nu}$.
The Yukawa interaction
\beq
\mathcal{L}_{{\rm R}_2}=  y_R^{ij} \, \overline{d}_{L\,i} \ell_{R\,j}\,{\rm R}_2^{(\sfrac{2}{3})}    +y_L^{ij} \overline{u}_{R\,i} \nu_{L\,j}\, {\rm R}_2^{(\sfrac{2}{3})} +\mathrm{h.c.}\,,
\label{eq:R2int}
\eeq
gives 
\beq
C_{S_L} (\mu_{\rm LQ})
&= 4 \, C_T(\mu_{\rm LQ}) 
= \dfrac{1}{4 \sqrt{2} G_F V_{cb}} \dfrac{y_{L}^{c \tau}\big{(}y_R^{b \tau}\big{)}^\ast}{m_{R_2}^2 }\,.
\label{eq:R2charged}
\eeq
In contrast to the above two LQ scenarios, the $\text{R}_2$ LQ does not generate $C_{V_L}$ but $C_{V_R}$. 
Thus we could expect solid predictions in polarization and related observables.
To generate $C_{V_R}$, indeed, a large mixing between two distinct $\text{R}_2$ LQ doublet is required to induce a proper electroweak symmetry breaking. 
See details in Refs.~\cite{Asadi:2019zja,Endo:2021lhi}.


\bibliographystyle{utphys28mod}

\bibliography{ref}

\end{document}